\documentclass[journal]{IEEEtran}

\usepackage{amssymb}
\usepackage{cite}

\usepackage{xcolor}
\usepackage[pdftex]{graphicx}

\usepackage{multirow}

\usepackage{amsmath}

\usepackage{algorithmic}

\usepackage[caption=false,font=footnotesize]{subfig}
\usepackage{url}

\begin{document}

\title{Orthogonal Transform based Generative Adversarial Network for Image Dehazing}
\author{Ahlad Kumar,~\IEEEmembership{Senior Member, IEEE,} Mantra Sanathra, Manish  Khare,~\IEEEmembership{Senior Member, IEEE,} and Vijeta Khare~\IEEEmembership{}

\thanks{Dr. Ahlad Kumar and Dr. Manish Khare are working as an Assistant Professor  at DA-IICT. (e-mail: ahlad\_kumar@daiict.ac.in, manish\_khare@daiict.ac.in)}
\thanks{
Mantra Sanathra is currently pursuing his Master's degree in Information and Communication Technology from DA-IICT, Gandhinagar, Gujarat (e-mail: mhsanathra@gmail.com) }
\thanks{Dr. Vijeta Khare is working an Assistant Professor at Adani Institute of Infrastructure Engineering (e-mail: vijeta.khare@aii.ac.in)}}

\maketitle

\begin{abstract}
Image dehazing has become one of the crucial preprocessing steps for any computer vision task. Most of the dehazing methods try to estimate the transmission map along with the atmospheric light to get the dehazed image in the image domain. In this paper, we propose a novel end-to-end architecture that directly estimates dehazed image in Krawtchouk transform domain. For this a customized Krawtchouk Convolution Layer (KCL) in the architecture is added. KCL is constructed using Krawtchouk basis functions which converts the image from the spatial domain to the Krawtchouk transform domain. Another convolution layer is added at the end of the architecture named as Inverse Krawtchouk Convolution Layer (IKCL) which converts the image back to the spatial domain from the transform domain. It has been observed that the haze is mainly present in lower frequencies of hazy images, wherein the Krawtchouk transform helps to analyze the high and low frequencies of the images separately. We have divided our architecture into two branches, the upper branch deals with the higher frequencies while the lower branch deals with the lower frequencies of the image. The lower branch is made deeper in terms of the layers as compared to the upper branch to address the haze present in the lower frequencies. Using the proposed Orthogonal Transform based Generative Adversarial Network (OTGAN) architecture for image dehazing, we were able to achieve competitive results when compared to the present state-of-the-art methods.

\end{abstract}


%
\IEEEpeerreviewmaketitle

\section{Introduction}
\begin{figure}[ht]
\centering
\subfloat[]{
  \includegraphics[width=0.45\linewidth]{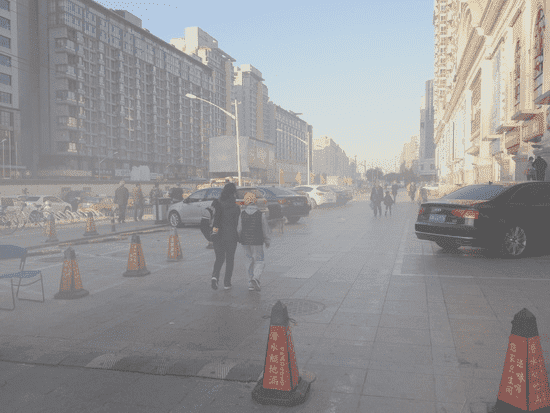} 
}
\subfloat[]{
  \includegraphics[width=0.45\linewidth]{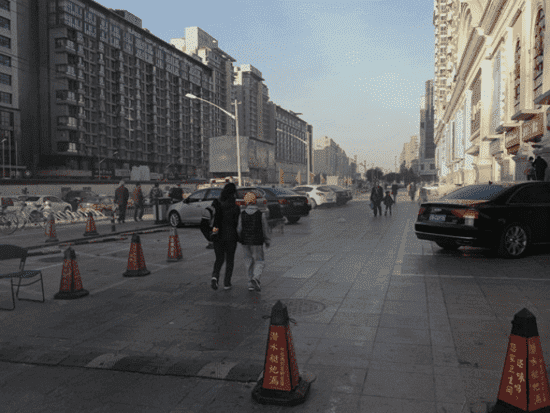}  
}
\caption{Example of Image Dehazing (a) Hazy Image (b) Clear Image}
\label{example}
\end{figure}

%
%
%
%
\IEEEPARstart{G}{enerally} it is difficult to capture a clear photo, especially in winter seasons. Some amount of fog or haze is present in the atmosphere and we do not have camera sensors that can directly remove this haze to overcome this problem. Haze is a natural phenomenon that degrades the quality of the image captured by the camera. This is due to fine particles like dust, water droplets, fog present in the atmosphere which absorbs and scatters the light. In order to address this problem, image dehazing is used to recover a haze-free image from a hazy image (Fig. \ref{example}). Computer vision tasks such as object detection \cite{objectdetection}, traffic surveillance, object tracking \cite{objecttracking} require a haze-free image to perform at their best potential. Thus haze removal becomes an essential pre-processing step for high-level computer vision tasks.
Earlier in \cite{chromatic,2002vision, depth,schechner2001instant,shwartz2006blind}, researchers have used multiple images of the same scene to recover the haze-free image. But still, it is not always possible to get multiple images of the same scene, which motivated them to perform image dehazing using a single image. Many methods were developed to address image dehazing task using a single image.

Image dehazing methods can be classified into two categories: (a) based on prior knowledge, (b) based on learning. The first one uses characteristic differences like brightness, contrast, saturation between hazy and haze-free images and utilizes this knowledge to obtain a haze-free image. But not all images show the same characteristics which lead to some artifacts (color distortion) that makes the dehazed image look unrealistic. On the other hand, learning-based methods extract these characteristics automatically using some learning model. 

In this paper, an orthogonal transform based Generative Addversarial network (OTGAN) is proposed for image dehazing.  The key aspects of the paper are mentioned below:
 \begin{itemize}
    \item GAN based deep learning architecture for image dehazing is introduced in orthogonal transform domain. Krawtchouk moments converts the images from spatial domain to Krawtchouk domain. The architecture is trained to find the difference between the Krawtchouk coefficients of hazy image and haze-free image. 
    \item Two custom convolution layers are designed consisting of Krawtchouk basis which is used to convert image in-between spatial domain and Krawtchouk domain; one of them is Krawtchouk Convolution Layer ($KCL$) used for forward transform and other Inverse Krawtchouk Convolution Layer ($IKCL$) for inverse transform. $KCL$ is kept fixed and non-trainable, while $IKCL$ is kept trainable for better adaptivity of the basis functions to the dataset.
    \item The proposed architecture has two branches; the upper branch consists of simple U-Net architecture, which deals with the high frequencies and the lower branch consist of pyramidal architecture that deals with the low frequencies present in the image
    \item Images used for training are transformed from  $RGB$ to $YCbCr$ color system, whereby only the $Y$ channel is passed through the architecture. 
    
\end{itemize}

The rest of the paper is structured as follows: Section \ref{rw} describes various methods used for image dehazing till now; Section \ref{fd} discusses image analysis in the frequency domain; Section \ref{km} provides an insight on using Krawtchouk moments and its role in image dehazing. The architecture of the proposed Orthogonal Transform based Generative Adversarial Network (OTGAN) for image dehazing is given in Section \ref{pm}. The details about various experiments carried out to analyse the performance of our proposed architecture is provided in Section \ref{expwork} along with the implementation details. Lastly, the failure cases and conclusion are mentioned in Sections. \ref{failcase} and \ref{conc} respectively.

\section{Related Work}
\label{rw}

\subsection{Haze formation formula}
\begin{figure}[h]
    \centering
    \includegraphics[width=\linewidth]{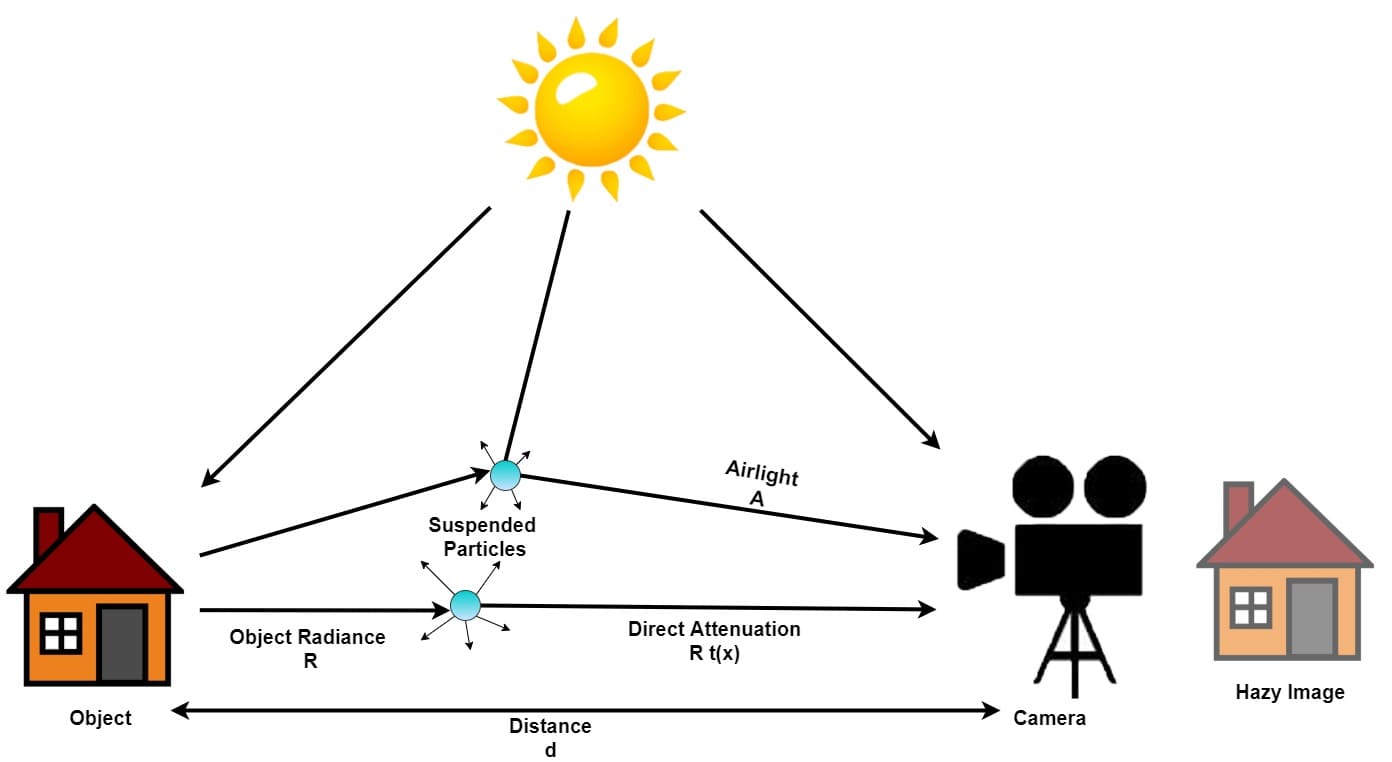}
    \caption{Haze Formation Model}
    \label{asm}
\end{figure}

 Fig. \ref{asm} shows the haze formation model. The atmospheric scattering model \cite{scattering} defines the haze formation model as
\begin{equation}
\label{model}
  I(x) = R(x)t(x) + A(1-t(x))  
\end{equation}
Here, $I$ stands for the image captured by the lens, $R$ stands for the haze-free image that we are trying to recover, $t$ stands for transmission map, which denotes the amount of light captured by the camera without any dispersion, and $A$ stands for the global atmospheric airlight. Transmission map $t$ is dependent upon the distance between camera lens and object and is calculated as
\begin{equation}
\label{trans}
 t(x) = e^{-\beta d} 
\end{equation}
where, $d$ represents the distance between the object and the camera lens. It can be seen from \eqref{trans} that the transmission map ($t$) is inversely proportional to $d$, so the objects near to the camera lens have less haze. This model is widely used by the researchers in estimating the clear images. The synthetic datasets can also be generated using the atmospheric scattering model by selecting a random value for the transmission map $t$ and random airlight $A$. These values are then used to generate hazy images from the clear images.

\subsection{Based on prior knowledge}
Earlier before the deep learning era, researchers used to find characteristics difference between brightness, contrast, saturation of the hazy and haze-free image and use this knowledge to estimate the transmission map $t$ and global atmospheric light $A$ and utilize \eqref{model} to get the clear image.

He \MakeLowercase{\textit{et al.}} \cite{dcp} introduced single image dehazing method using Dark Channel Prior (DCP), which is based on the observation that, outside haze-free images have some local regions whose intensity value is very low (near to zero) for at least one of the color channel. They used this observation along with the atmospheric scattering model to directly estimate the clear image from the estimated transmission map and atmospheric light as follows

\begin{equation}
\label{dcp}
 R = \frac{I-A}{max(t,t_{0})}
\end{equation}
Here, $t_{0}$ denotes the lower bound of the transmission map and $A$ denotes the global atmospheric airlight.This method was not able to produce good results for regions that are similar to airlight. The observations made in DCP were used by many researchers in their work. An efficient image dehazing with Boundary Constraint and Contextual Regularization (BCCR) \cite{bccr} by Meng \MakeLowercase{\textit{et al.}}, proposed a boundary constraint on transmission function and utilized it to estimate the transmission map. In Non-local Image Dehazing (NLD) \cite{nld} by Berman \MakeLowercase{\textit{et al.}}, utilized a non-local prior knowledge for image dehazing. They observed that only a few hundred distinct colors are required to represent a haze-free image, which is tightly clustered in $RGB$ space. These color clusters behaves differently for hazy and haze-free images. The color cluster present in the hazy image becomes the haze line in the haze-free image and this knowledge is used to estimate the transmission map and further used in recovering the haze-free image.    

Many methods are introduced for image dehazing in the spatial domain, Liu \emph{et al.} introduced a novel approach to dehaze image in the frequency domain \cite{liu2017efficient}. They used multi-scale wavelet decomposition \cite{wavelet} to convert images from spatial domain to frequency domain. It was observed that haze is present in the low frequency content of the image; wavelet decomposition produces four different sub-images where one image contains low-frequency content while the other contains high-frequency content specifically they contain horizontal, vertical and diagonal details of the image.  The authors proposed Open Dark Channel Model (ODCM) for removing haze from low-frequency part and the transmission value obtained from ODCM is used to reduce the noise from high-frequency part of the image and finally, haze-free image is obtained from wavelet decomposition. 

Prior based methods are fast as they do not require any training, but they work on the assumptions made by the authors such as dark channel, color attenuation which are not true for all kinds of images. Even though these methods can remove the haze but the clear image does not look realistic due to some color distortion and oversaturation. This can be solved using some optimization but each image requires a different type of optimization which is not feasible. To overcome these problems, researchers started using learning-based methods which will be discussed next.

\subsection{Based on learning}
Zhu \MakeLowercase{\textit{et al.}} proposed Color Attenuation Prior (CAP) \cite{cap} which uses prior knowledge along with linear learning model to estimate the scene depth. CAP is based on the fact that the difference between the saturation and brightness varies for hazy and haze-free image and it is directly proportional to the depth map of the image. So the authors have used supervised linear learning model to estimate the depth map. Cai \MakeLowercase{\textit{et al.}} proposed a CNN based DehazeNet \cite{dehazenet}, architecture using different convolution layers stacked together to estimate the transmission map and further recover the haze-free image; they also introduced BReLU  for accurate restoration of the image. MSCNN\cite{mscnn} is CNN based architecture that uses two different branches for estimating transmission maps, one of the branches estimates at coarse-scale and the other at the fine-scale. 

Most of the methods used learning methods to estimate the transmission map and simply use prior knowledge to obtain the global atmospheric airlight. Shin \MakeLowercase{\textit{et al.}} \cite{rrc} proposed a novel optimization framework that integrates radiance and reflectance components along with structure-guided $l_0$ norm for further refinement. This reflectance map is used to estimate the transmission map which is further used for image dehazing.
In All-in-one Dehazing Network(AOD-Net) \cite{aod}, Li \MakeLowercase{\textit{et al.}} modified the atmospheric scattering model by combining the transmission map and  airlight into one single term. Using lightweight CNN, the clear image is estimated directly instead of estimating the transmission map first. Li \MakeLowercase{\textit{et al.}} proposed PDR-Net \cite{pdrnet}, which uses CNN to reconstruct dehazed image and further a network is used to enhance the color and contrast properties of the dehazed image.   Lin \MakeLowercase{\textit{et al.}} \cite{msaff-net} proposed end-to-end attention based lightweight model MSAFF-Net which uses a channel and multiscale spatial attention module, for determining the regions with  haze-related features. Zhang \MakeLowercase{\textit{et al.}} proposed a Densely Connected Pyramid Dehazing Network (DCPDN) \cite{dcpdn} which estimates the transmission map and airlight jointly to obtain the dehazed image. Authors proposed an encoder-decoder based on the densely connected network along with pyramid pooling to estimate the transmission map and U-Net\cite{unet} is used to estimate the airlight. Discriminator based on GAN\cite{gan} framework is used to decide whether the estimated image is real or fake. 

Learning-based methods achieved accurate results but a large amount of data is required during the training process. It is difficult to get ground truth images for real-world hazy images so synthetic datasets are used during training. Because of this, learning-based methods are not able to dehaze real-world images completely which opens a space for further research.

\section{Motivation}
\label{fd}
In \cite{liu2017efficient} authors performed image dehazing in the frequency domain instead of the spatial domain. Wavelet transform is used to convert the image from the spatial domain to the frequency domain. It has been observed that the hazy images have more content in the low-frequency spectrum while haze-free images have less content in the low-frequency spectrum. One of the reasons for this could be that the haze-free images are sharper and contain more edges as compared to hazy images. From this important observation, it is concluded that the haze is generally present in the lower frequency spectrum. Motivated by this observation, Krawtchouk moments are used to transform images from spatial domain to orthogonal domain in this paper. The details about Krawtchouk moments and its analysis on hazy images is discussed next.
 
\section{Krawtchouk Moments}
Krawtchouk moment is widely used in the area of pattern recognition \cite{priyal2013robust,rahman2016selection}. They are well suited as pattern features in the analysis of two-dimensional images and can be used for image dehazing. In this section, a brief review about the definition of Krawtchouk moment is discussed followed by its role in the area of image dehazing.
\label{km}
\subsection{Computation of Krawtchouk Moments}
Image analysis using Krawtchouk moments introduced a new set of orthogonal moments based on the discrete classical Krawtchouk polynomials\cite{krawtchouk1929interpolation} associated with the binomial distribution.  Krawtchouk moments of order $(m+n)$ for an image $g(x,y)$ is given as \cite{yap2003image}
\begin{equation}
   Q_{nm} = \sum_{x=0}^{N-1}\sum_{y=0}^{N-1}\bar{K}_{n}(x;p_{1},N-1)\bar{K}_{m}(y;p_{2},N-1)g(x,y)
   \label{eq1x}
\end{equation}
with $n=0,1,...,N-1$; $m=0,1,...,N-1$; $g(x,y)$ is image with size of $N \times N$, $\bar{K}_m$ and $\bar{K}_n$ is set of weighted Krawtchouk polynomials, given as 
\begin{equation}
    \bar{K}_n =(x;p,N) = K_n(x;p,N)\sqrt{\frac{w(x;p,N)}{\rho(n;p,N)}}
\end{equation}
where
\begin{equation}
    w(x;p,N)= \binom{N}{x} p^x(1-p)^(N-x)
\end{equation} 
and,
\begin{equation}
    \rho(n;p,N)= (-1)^n \left(  \frac{1-p}{p}\right)^{n}\frac{n!}{(-N)_n}
\end{equation}
and $K_n(x;p,N)$ is $n$-th order classical Krawtchouk polynomial defined as
\begin{equation}
    K_{n}(x;p,N) = \sum_{k=0}^{N}a_{k,n,p}x^{k} = _2F_{1} \Bigg(-n,-x;-N;\frac{1}{p} \Bigg).
\end{equation}
where $x,n=0,1,2,....,N,N>0,p\in(0,1)$. The hypergeometric function $_2F_1$ is defined as
\begin{equation}
    _2F_1(a,b;c;z)=\sum_{k=0}^{\infty}\frac{(a)_k(b)_k}{(c)_k}\frac{z^k}{k!}
\end{equation}
where $(a)_k$ is the Pochhammer symbol given by
\begin{equation}
    (a)_k =a(a+1)\dots(a+k-1)=\frac{\Gamma(a+k)}{\Gamma(a)}
\end{equation}

The image can be reconstructed from Krawtchouk moments using the following equation as
\begin{equation}
    g(x,y) = \sum_{x=0}^{N-1} \sum_{y=0}^{N-1}Q_{nm}\bar{K}_n(x;p_1,N-1)\;\bar{K}_m(y;p_2,N-1)
    \label{eq2}
\end{equation}

\subsection{Representation in Matrix Form}
Krawtchouk moment given in (\ref{eq1x}) can also be implemented in matrix format. The set of Krawtchouk moments upto order $(m + n)$ in matrix form is given as
\begin{equation}
    \textbf{Q} = \textbf{K}_{2}\textbf{G}\textbf{K}_1^T
\end{equation}
where $\textbf{G}$ is the image matrix, $K_1$ and $K_2$ are Krawtchouk polynomial matrix derived from  matrix $\textbf{K}_v$ with $v$=1,2 as follows 


\begin{equation}
    \textbf{K}_{v} =
    \begin{bmatrix}\bar{K}_0(0;p_v,N-1) & \dotsm & \bar{K}_0(N-1;p_v,N-1)  \\  \vdots & \ddots & \vdots \\\bar{K}_{N-1}(0;p_v,N-1) & \dots & \bar{K}_{N-1}(N-1;p_v,N-1)
    \end{bmatrix}
\end{equation}
The inverse transformation given in (\ref{eq2}) can be represented in the matrix form as

\begin{equation}
    \textbf{G} = \textbf{K}_{2}^{T}\textbf{Q}\textbf{K}_1
\end{equation}

\subsection{Basis function of Krawtchouk Moments}
\begin{figure}[h]
\centering
\subfloat[ \label{basis}]{\includegraphics[width=0.45\linewidth ]{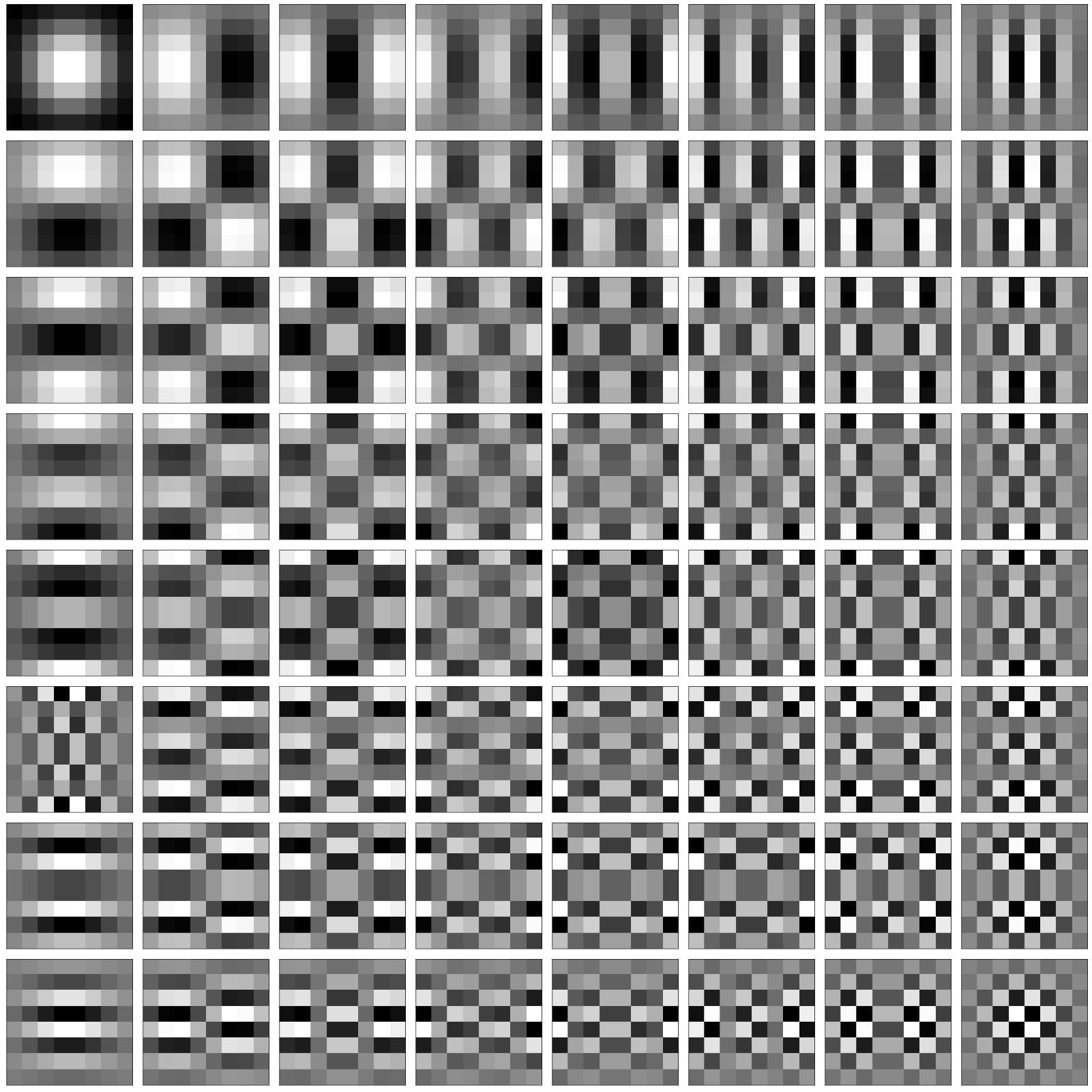}}
\subfloat[ \label{zigzag}]{\includegraphics[width=0.45\linewidth ]{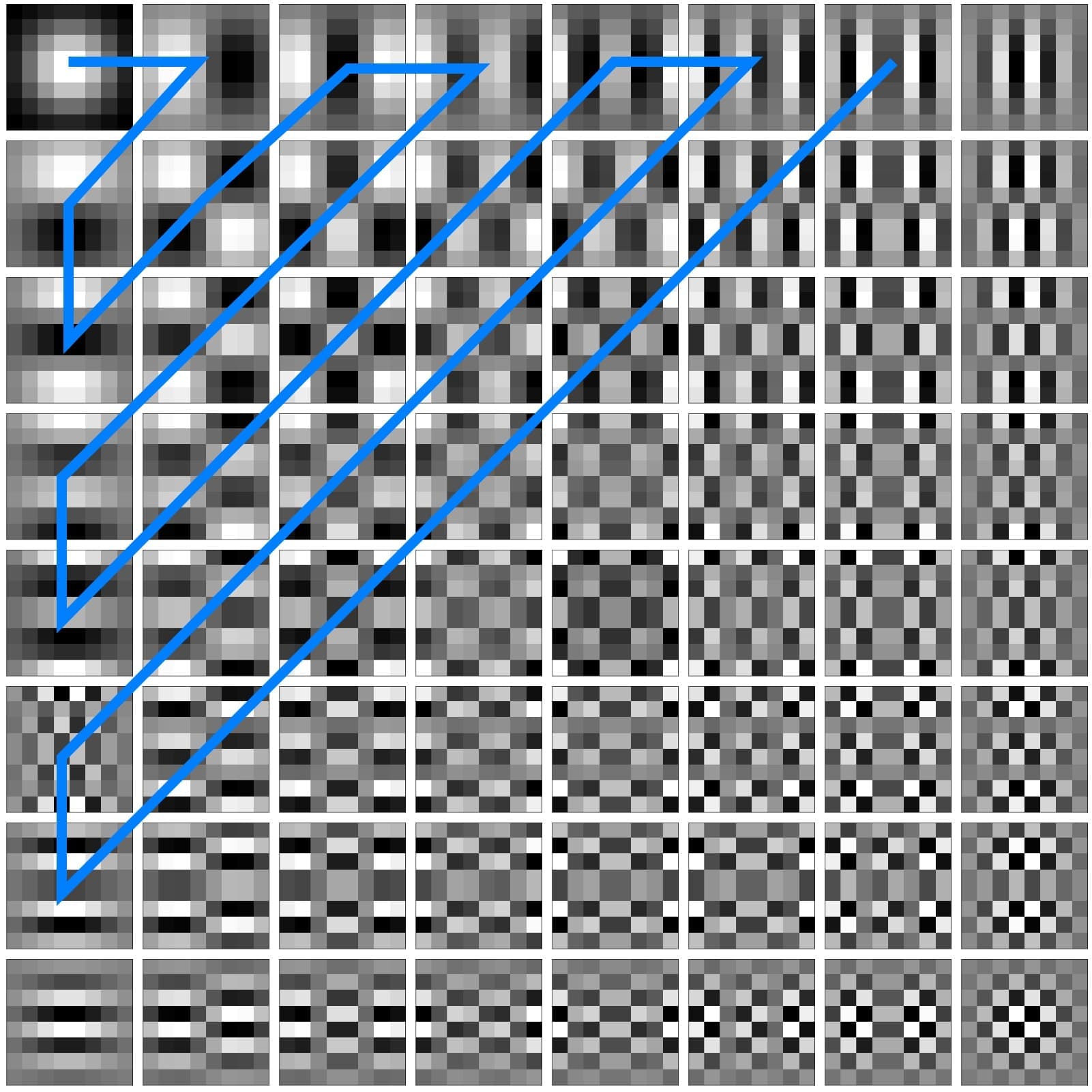}
}

\caption{(a) Basis function of Krawtchouk moments, (b) Zig-zag ordering of basis functions}

\end{figure}

Krawtchouk moments of an image can be interpreted as the projection of the image on the basis functions,$w_{i,j}$ which is given as
\begin{equation}
    w_{i,j} = [k_{i}]^{T}[k_{j}] 
\end{equation}

where
\begin{equation}
    k_i = \left[\bar{K}_i(0;p,N-1),\;\ldots\;,\bar{K}_i(N-1;p,N-1) \right] 
\end{equation}
and
\begin{equation}
    k_j = \left[\bar{K}_j(0;p,N-1),\;\ldots\;,\bar{K}_j(N-1;p,N-1) \right] 
\end{equation}
with $i=0,1,..,N-1$ and $j=0,1,...,N-1$.
The basis function $w_{i,j}$ is shown in Fig. \ref{basis}. The value of $N$ amd $p$ is taken as 8 and $p$ respectively. Krawtchouk moments of an image also provides a correlation between image $\textbf{F}$ and basis function i.e., the value of the coefficient is higher if there is a strong similarity between the basis function and the image content and vice versa.

Krawtchouk basis functions are used as filters in the proposed architecture. Inspired from the JPEG (Joint Photographic Experts Group) compression method \cite{jpeg} basis are rearranged in  the zig-zag manner as shown in Fig. \ref{zigzag}.

\begin{figure*}[!ht]
\subfloat[]{
  \includegraphics[width=0.32\linewidth]{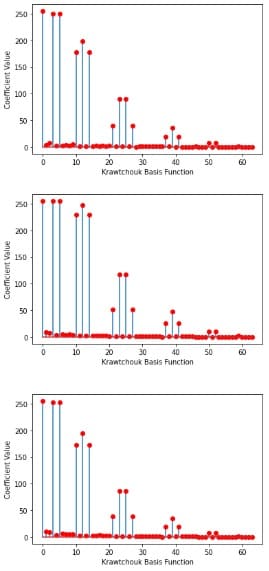} }
\subfloat[]{
  \includegraphics[width=0.32\linewidth]{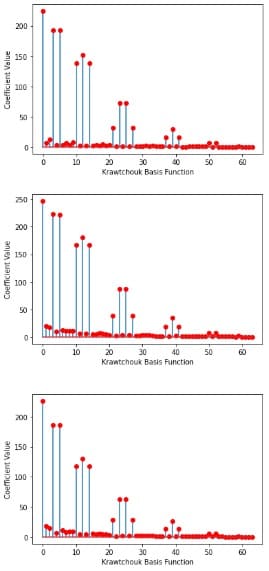}   }
\subfloat[]{  
  
  \includegraphics[width=0.32\linewidth]{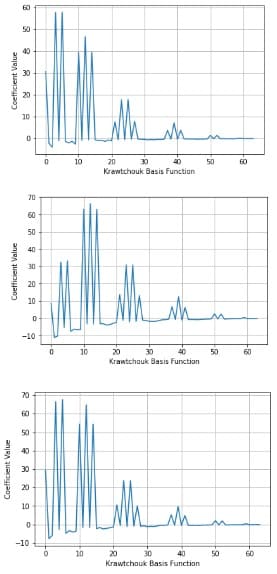} 
}
\caption{(a) Krawtchouk coefficients for hazy image (b): Krawtchouk coefficients for clear image (c) Difference in the coefficients of hazy and clear image }
\label{kcoff}
\end{figure*}

\begin{figure*}[!ht]
\centering
\includegraphics[width=\linewidth ]{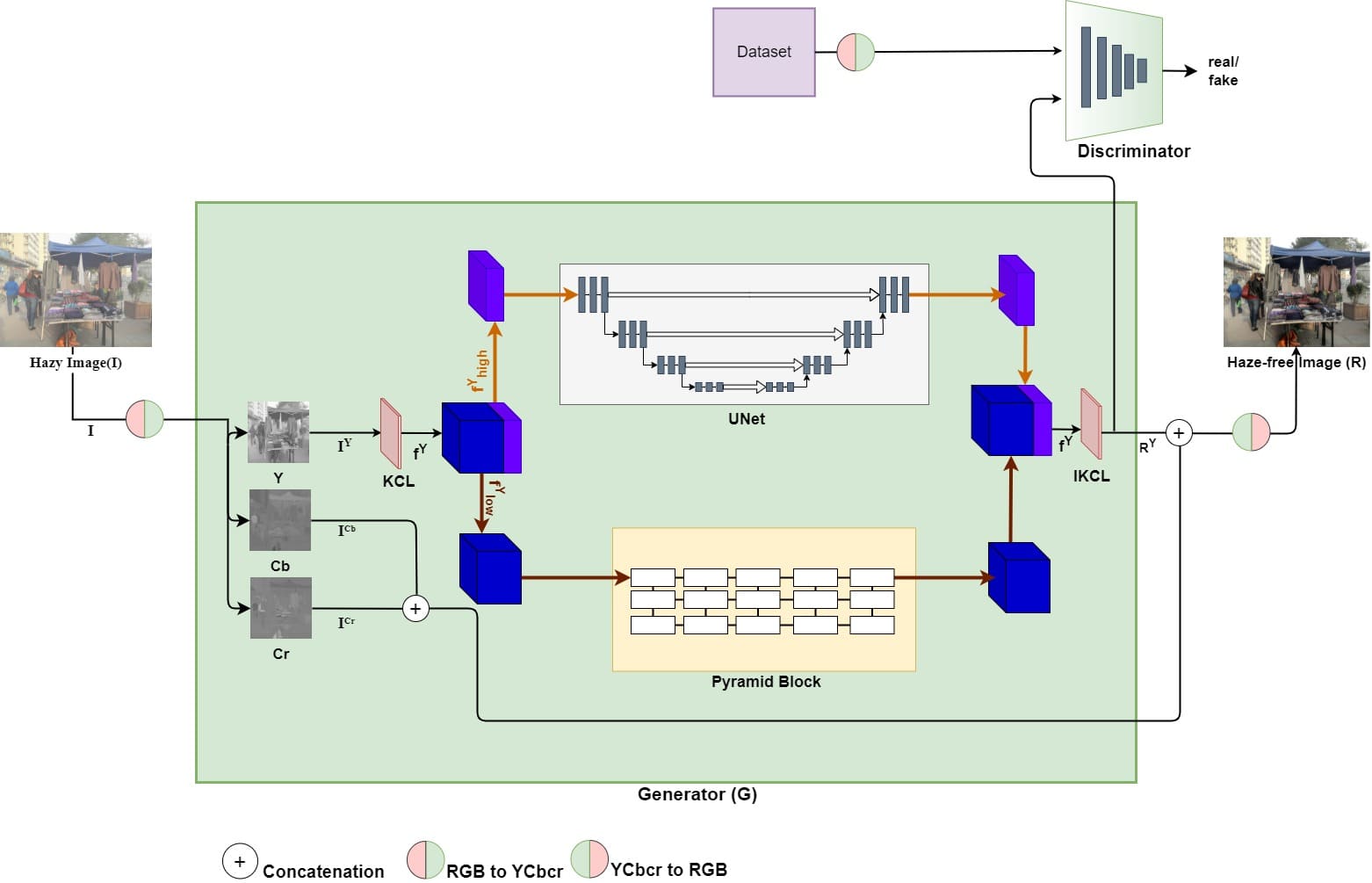}
\caption{Architecture of proposed model, hazy image is provided as input to the model and further, it is converted into $YCbCr$ mode and only $I^Y$ channel is passed to model, model is divided in two branches $f^Y_{low}$(shown with brown arrow) and $f^Y_{high}$(shown with orange arrow), at the end two branches are combined to get haze-free image.}
\label{architecture}
\end{figure*}

We have used 64 such basis functions and represented them using $w_i$ where $i=0,1,...,63$. Zig-Zag ordering arranges the basis functions in increasing order of frequency, i.e., frequency component increases from low to high with the increase in index $i$.
Average values of coefficients generated from the convolution of basis functions with three different hazy and clear images is shown in Fig. \ref{kcoff}. Here Fig. \ref{kcoff}(a)-(b) shows coefficients of three different hazy and clear image of the same scene whereas Fig. \ref{kcoff}(c) shows the difference between these coefficients. It can be seen from Fig. \ref{kcoff}(c) that there is a significant loss of Krawtchouk coefficients in basis functions with lower frequency components. Thus, in the Krawtchouk domain, the task of dehazing reduces to recovering the low-frequency Krawtchouk coefficients of a clear image from its corresponding hazy image. This observation is used in the proposed architecture discussed in the next section.

\section{Proposed method}
\label{pm}

In this section, the proposed architecture, shown in Fig. \ref{architecture} is discussed in details. It consists of 8 blocks: (1) $RGB$ to $YCbCr$ (2) Krawtchouk Convolution Layer ($KCL$) (3) Frequency Cube (4) Pyramidal block for lower frequency (5) U-Net block for higher frequency (6) Inverse Krawtchouk Convolution Layer ($IKCL$) (7) Discriminator (8) $YCbCr$ to $RGB$ .The details of the mentioned blocks are discussed next.

\subsection{Architecture Structure}
\subsubsection{Colour Space Transformation: RGB to YCbCr}
Whenever we capture any image, it needs to be stored in the electronic devices such as computers which only understand numbers. Hence, some rules need to be followed while storing the images in the memory. The color space defines this set of rules. Generally, $RGB$ color space is used which uses Red-Green-Blue color components of an image to represent any image. The $YCbCr$ is another type of color space which represents the image using $Y$, $Cb$, and $Cr$ components of the image. The $Y$ component represents the Luma (brightness) component of the image, $Cb$ and $Cr$ represent the blue and red components related to the chroma component. 

\begin{figure}[!ht]
\centering
\includegraphics[width=\linewidth ]{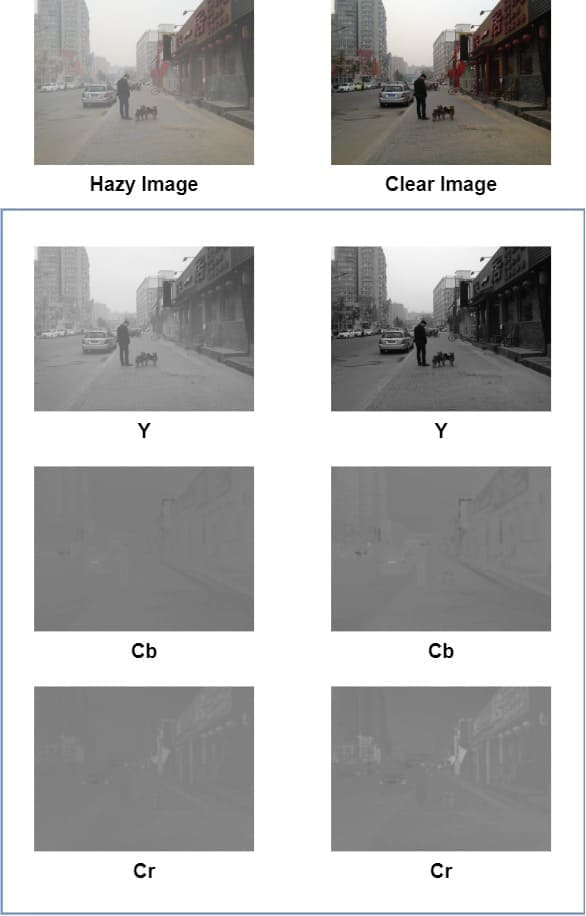}
\caption{Image analysis in $YCbCr$ color space}
\label{ycbcr}
\end{figure}

Fig. \ref{ycbcr} shows the hazy image along with its corresponding haze-free image in $YCbCr$ color space. It can be seen that the haze component is mainly present in the $Y$ channel of the image. Hence, it plays an important role as compared to $Cb$ and $Cr$ components. $Y$ channel of the hazy and haze-free image shows significant difference while the $Cb$ and $Cr$ channels do not have a significant difference. From this crucial observation, we decided to only use the $Y$ channel for estimating the haze-free image and not changing the $Cb$ and $Cr$ channels. Considering this fact, first the $RGB$ image is converted to $YCbCr$ color mode so that the hazy and haze-free image pairs can be compared in $YCbCr$ space. Next, only the $Y$ channel is passed through the proposed architecture instead of all channels. The different channel are represented as $I^{Y}, I^{Cb} \;and\;I^{Cr}$ and are shown in Fig.\ref{architecture}.

\subsubsection{Krawcthouk Convolution Layer (KCL)} This layer transforms images to the Krawtchouk moments domain (orthogonal domain) from the spatial domain. Krawchouk basis function $w_i$ of size $8 \times 8$ are treated as the filters. There are a total of 64 such filters ($w_i$) arranged in a zig-zag manner (Fig. \ref{zigzag}). The $KCL$ layer consist of 64 features maps $f_{i}$ created by performing convolution operation of $w_i$ with \textbf{$I^Y$}  as follows
\begin{equation}
\label{kcl}
f_{i} =  w_i \circledast I^{Y} \quad \forall i \in \{ 0,1,2,...,63 \} 
\end{equation}
 Here, $\circledast$ represents convolution operation in which stride \textcolor{red}{$S$} is kept 1 and padding is kept as \emph{same} for retaining the size of the image. The $KCL$ layer is kept fixed and non-trainable during the training phase and its functionality can be compactly represented as follows
\begin{equation}
    f^{Y} = KCL(f_i)   
    \label{eq3}
\end{equation}
Here, $f^{Y}$ represent the frequency cube containing all the feature maps ranging from 0 to 63. The details about the frequency cube is discussed next.
\subsubsection{Frequency Cube $(f^{Y})$} \label{tar}
\begin{figure}[h]
    \centering
    \includegraphics[width=\linewidth]{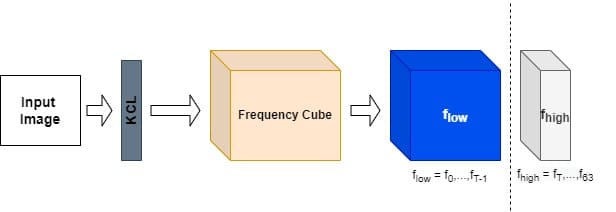}
    \caption{Frequency Partitioning}
    \label{cube_split}
\end{figure}

The feature maps obtained from (\ref{eq3}) are used to form a frequency cube $f^{Y}$. This cube is ordered in the increasing order of the frequency content. The cube is split into two parts from a particular point \emph{T}. Two parts are denoted as $f^{Y}_{low} = f^{Y}_0,...f^{Y}_{T-1}$
and $f^{Y}_{high} = f^{Y}_{T},f^{Y}_{T+1},...,f^{Y}_{63}$. The optimal value of the split point \emph{T} is obtained experimentally and its value is found to be 60. The details about how to select this value is discussed in the experimental section. The process of partitioning is shown in Fig. \ref{cube_split}. The partitioned cubes $f^{Y}_{low}$ and $f^{Y}_{high}$ are processed separately. As discussed in Fig. \ref{kcoff}, the Krawtchouk coefficients have a substantial loss in lower frequencies compared to high frequencies. So, $f^{Y}_{low}$ block requires complex architecture to recover the haze-free image from the hazy image, while simple architecture can be used for $f^{Y}_{high}$ block. Next, we will discuss the network architecture for dealing with both these frequency blocks $f^{Y}_{low}$ and $f^{Y}_{high}$ respectively. 

 \subsubsection{Architecture for $f^{Y}_{low}$}
\begin{figure}[h]
    \centering
    \includegraphics[width=\linewidth]{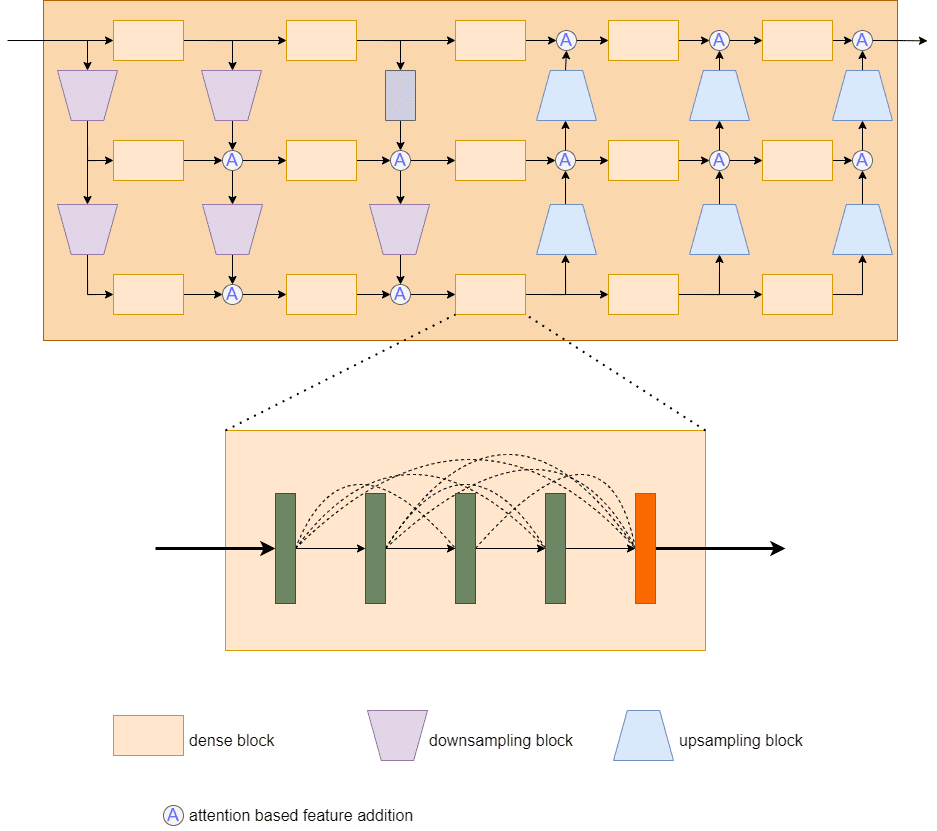}
    \caption{Architecture for lower branch $(f^{Y}_{low})$}
    \label{f_low}
\end{figure}

Taking motivation from \cite{gridnet}, we have used a similar kind of structure for the lower part of the architecture. The detailed structure of the lower branch of the proposed architecture is shown in Fig. \ref{f_low}. The frequency cube  $f^{Y}_{low}$ obtained from frequency partitioning is sent as an input to this network which consists of six columns and three rows. The first three columns consist of down-sampling blocks and the remaining three consist of the up-sampling block.  The up-sampling block increases the number of feature maps by a factor of two and the down-sampling block decreases the number of feature maps by the factor of two. Due to this, each row which contains five dense blocks, performs an operation on a different scale while keeping the number of feature maps the same. As the feature maps of different scales have different importance, an attention mechanism is also incorporated. 

Next, we will discuss the structure of the dense block shown in Fig. \ref{f_low} which is used in the architecture for  $f^{Y}_{low}$. Each dense block consists of five convolution layers in which first four layers increases the feature map and has a skip connection with previous layers. The last layer fuses all these feature maps together such that number of feature map is equal to the number of input feature map. 


\subsubsection{Architecture for $f^{Y}_{high}$}
The higher frequency cube $f^{Y}_{high}$ obtained from the frequency partitioning is sent to the upper part of the architecture. As higher frequencies do not show a substantial loss in Krawtchouk coefficients, therefore a simple UNet\cite{unet} structure is used for recovering higher frequency coefficients. The UNet structure used in the proposed architecture contains four encoders and decoders blocks. Each encoder block is constructed by stacking up convolution, batch-normalization and deconvolution blocks together. The size of the kernel is kept $128 \times 128$ for starting encoder block and is decreased by factor 2 for the successive encoder blocks; which is then increased by a factor of 2 for the successive decoder blocks.

\subsubsection{Inverse Krawtchouk Convolution Layer (IKCL)} 
\label{bn}
The outputs from the lower and upper branch of the architecture are combined at the end. As the image is in Krawtchouk moment domain, it needs to be transformed into the spatial domain. The $IKCL$ layer consists of a convolution layer that converts the image from the Krawtchouk moment domain to the spatial domain. The weights of the kernel are kept trainable during the training phase for providing better adaptivity of the basis functions to the dataset. This operation can be represented as follows
\begin{equation}
    R^{Y} = IKCL(f^{Y})
\end{equation}
where, $R^{Y}$ represents the image generated by the proposed architecture.
\subsubsection{Discriminator}
The generator and discriminator based GAN\cite{gan} framework is used for image dehazing. Hazy image is passed through the generator (orange box in Fig. \ref{architecture}), which directly estimates the haze-free image. The $Y$ channel of the generated  image is passed through the discriminator along with the $Y$ channel of the ground truth image. The discriminator is trained to decide whether the generated image is real or fake. The task of the generator is to produce a haze-free image that is indistinguishable from the ground truth. Discriminator and generator are not trained at the same time. The weights of discriminator are kept fixed during training of the generator, and during training of discriminator, the weights of the generator are kept fixed.

\subsubsection{Colour Space Transformation: YCbCr to RGB}
The image ($R^{Y}$) generated from the proposed architecture is combined with the $I^{Cb}$ and $I^{Cr}$ channels of the input image to get a haze-free image $R$ which is finally transformed from $YCbCr$ color-space to $RGB$ space for visualization.

\section{Experimental Work}

\label{expwork}


In this section, experiments are performed to verify the working of the proposed architecture and compared the results with the state-of-the-art methods. Quantitative and qualitative experiments are carried out on synthetic images as well as the real-world images having no ground truth.

\subsection{Datasets}
\begin{figure}[!h]
    \centering
    \includegraphics[width=\linewidth]{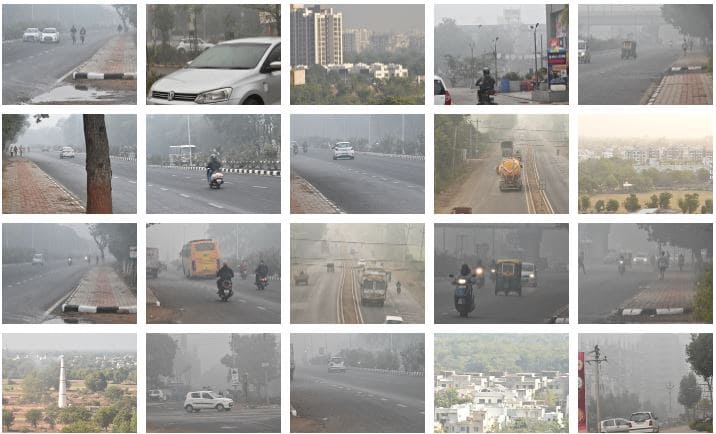}
    \caption{Real-world dataset of hazy images}
    \label{rwdata}
\end{figure}

Image dehazing is an ill-posed problem, and it is difficult to get a large number of hazy images along with its haze-free image. Most of the dehazing methods use synthetic datasets for training their models. For creating synthetic training datasets, a depth map of haze-free images is obtained either from the existing datasets or by estimating the depth map, and then using \eqref{model}, hazy image is generated. We have used RESIDE (REalistic Single Image DEhazing) \cite{reside} dataset which is a large scale synthetic dataset containing both outdoor (OTS) and indoor (ITS) hazy images along with its clear images. It is widely used for the training and testing of different dehazing algorithms. We have trained our model using the Outdoor Training Set (OTS) of RESIDE and testing is done on SOTS of RESIDE. The SOTS dataset contains 1000 pairs of hazy and clear images of 500 outdoor and 500 indoor scenes, generated in the same way as training data is generated. We also tested our model on the HSTS dataset of RESIDE which contains synthetic hazy image along with real-world images. Moreover, to validate the performance of the proposed architecture on real world dataset, we have created our own dataset of 200 real world hazy images. Some of the images from the dataset are shown in Fig. \ref{rwdata}.

\subsection{Loss Functions}
The selection of loss function plays an important role in training deep learning-based model. It has been observed through experiments that by simply using mean-square error (MSE) loss is not helpful as it does not perform well with the outliers. Hence, in this paper we have used weighted sum of three different types of losses, the details of which are as follows

\subsubsection{VGG Loss}
If we consider any deep neural-based image classification network, in the first few layers of the network, the feature maps obtained from the convolution layer generally contains the edges present in the image. These feature maps can be used as loss functions to find a difference between the estimated and ground-truth clear image. We have used a pre-trained VGG16 model \cite{vgg16} trained on ImageNet \cite{imagenet} as the loss network. The feature maps of the last layer of the first three stages are used for defining the VGG loss as follows
\begin{equation}
    \label{vggloss}
    L_{vgg} = \sum_{i = 1}^{3}\frac{1}{Ch_{i} M_{i} N_{i}}  \left\| \phi_{i}(\hat{R})-\phi_{i}(R) \right\| ^{2}_{2}
\end{equation}
where $Ch$ represents the channel, $M$ and $N$ represents the size of the image, \textbf{i} represents the stage of the VGG16 network, $\phi_{i} (\hat{R}$) and $\phi_{i}(R)$, represents the features maps of the VGG16 network.
Here, $\hat{R}$ and $R$ represents the estimated and the ground truth images respectively.

\subsubsection{Smooth $L_{1}$ Loss}
The \emph{$L_1$} loss is less sensitive to outliers as compared to MSE loss and is able to prevent possible gradient explosion \cite{l1loss}. Let $\hat{R}_i$ and ${R}_i$ represents the dehazed and the original image at pixel $i$ and $N$ is the total number of pixels. The smooth $L^{(s)}_1$ loss  \cite{l1loss} can be calculated as follows
\begin{equation}
    \label{l1}
      L^{(s)}_1(\hat{R},R) = \frac{1}{N} \sum_{i=1}^{N}\xi (\hat{R}-R)
\end{equation}   
where,

\begin{equation}
\xi (\hat{R}-R) = \xi(l) = \begin{cases}
0.5(l)^2 & if \quad |l| < 1 \\
|l|-0.5 & otherwise
\end{cases}
\end{equation}
\subsubsection{GAN Loss}
We have used GAN based architecture to determine whether the generated haze-free is real or fake. Discriminator $D$ tries to distinguish between real and fake images while generator $G$ is trained to produce haze-free images such that the discriminator is not able to differentiate between real and fake images. Let $G(I)$ denote the haze-free image generated by the generator, and $R$ indicates the real haze-free image from the dataset. The GAN loss can be calculated as follows
\begin{equation}
    L_{GAN} = \underset{G}{min}\; \underset{D}{max}\; \mathbb{E}[ R \;log(D(R))] \\
    + \mathbb{E}[I \;log(1-D(G(I)))]
\end{equation}

The total loss of the proposed model is obtained as a weighted sum of $L_1 ,L_{vgg}$ and $L_{MSE}$  as follows
\begin{equation}
    L = \lambda_1L_{vgg} + \lambda_2L_{1}^{(s)} +\lambda_3L_{MSE} +\lambda_4L_{GAN}
\end{equation}
Here, $\lambda_1,\lambda_2$, $\lambda_3$ and $\lambda_4$ are regularization parameters of the loss function.

\subsection{Implementation Details}
As the size of the model is large, training the model with a full image requires high computational power and will also take more time. Therefore, we have randomly selected patches of size $128 \times 128$ from hazy images and select their corresponding patch from the clear image. We have used Adam \cite{kingma2014adam} optimizer for fast learning with a batch size of 15. The learning rate is kept at 0.001. Inspired from \cite{ryf}, the training images are converted to $YCbCr$ color mode from $RGB$, and only the $Y$ (brightness) component is passed to the architecture and the remaining $Cb$ and $Cr$ channels are directly passed to the end of the architecture where is combined with the $Y$ channel of the clear image obtained from the architecture. For the loss function the values of parameters are taken as: $\lambda_1 = 0.5, \lambda_2 = 1$, $ \lambda_3=0.04$ and $ \lambda_4=0.05$. The model is trained for 20 epochs on NVIDIA RTX 3600.

\subsection{Optimization of IKCL}

\begin{figure}[h]
    \centering
    \includegraphics[width=\linewidth]{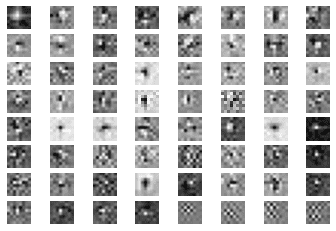}
    \caption{Optimized filters of $IKCL$}
    \label{ikcl_fig}
\end{figure}

Fig. \ref{ikcl_fig} shows the optimized basis functions of $IKCL$ layer after the training process of model is completed. As mentioned earlier in Section. \ref{bn}, $IKCL$ layer is kept trainable during the training phase for providing better adaptivity of the basis functions to the dataset which can be seen from the figure.

\begin{figure*}[ht]
\captionsetup{justification=centering}
\resizebox{\textwidth}{!}{
\subfloat[Hazy image]{
  \includegraphics[width=0.08\linewidth]{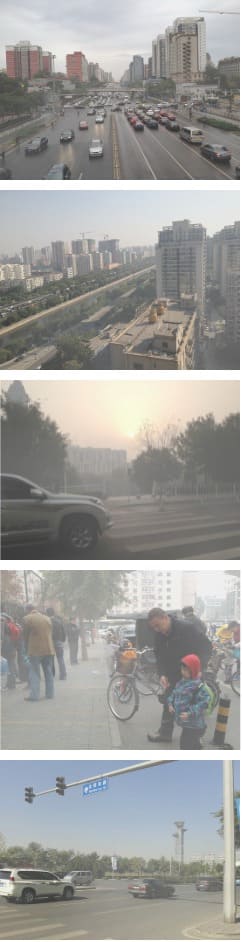}  
}\subfloat[DCP]{
  \includegraphics[width=0.08\linewidth]{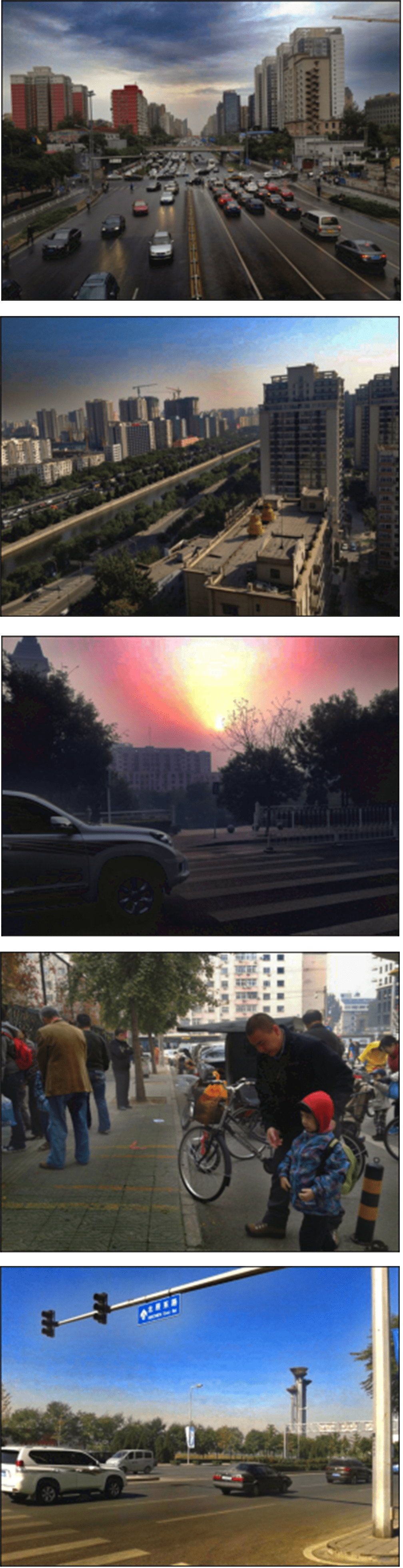}  
}
\subfloat[CAP \label{cap}]{
  \includegraphics[width=0.08\linewidth]{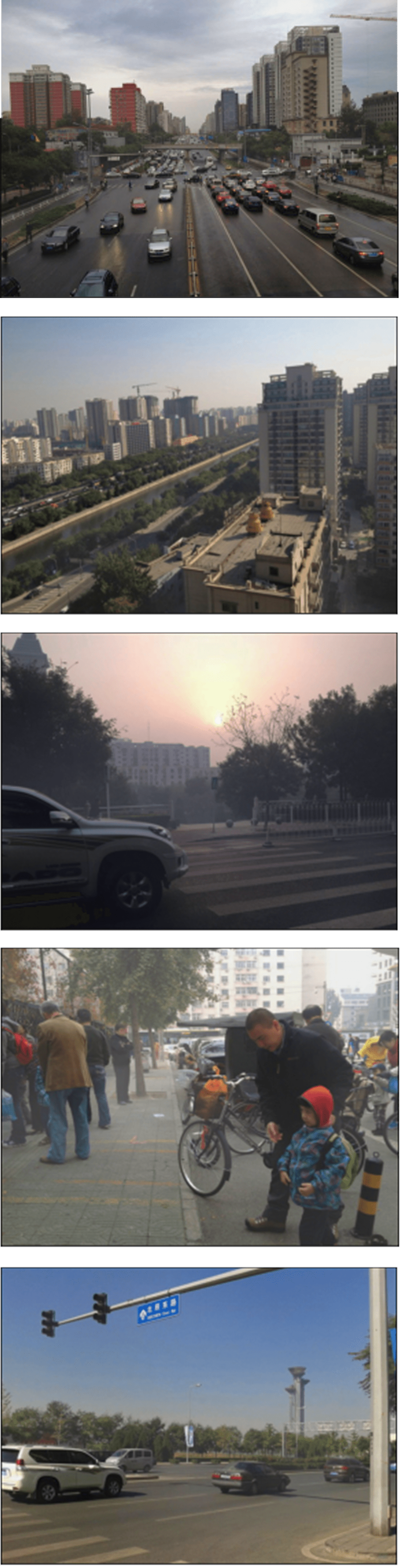}  
}
\subfloat[NLD]{
  \includegraphics[width=0.08\linewidth]{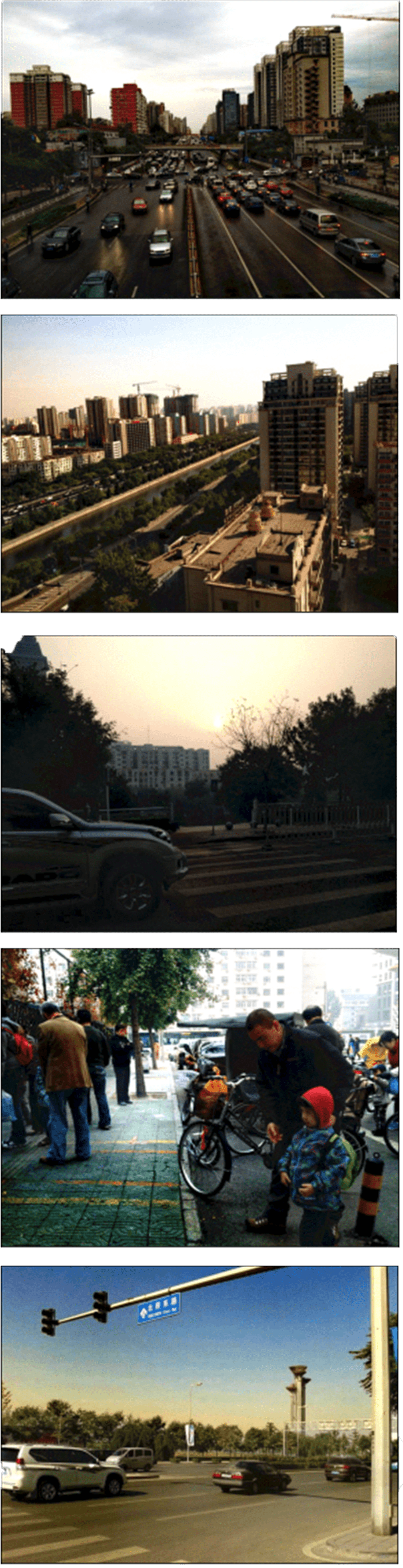}  
}
\subfloat[BCCR \label{bccr}]{
  \includegraphics[width=0.08\linewidth]{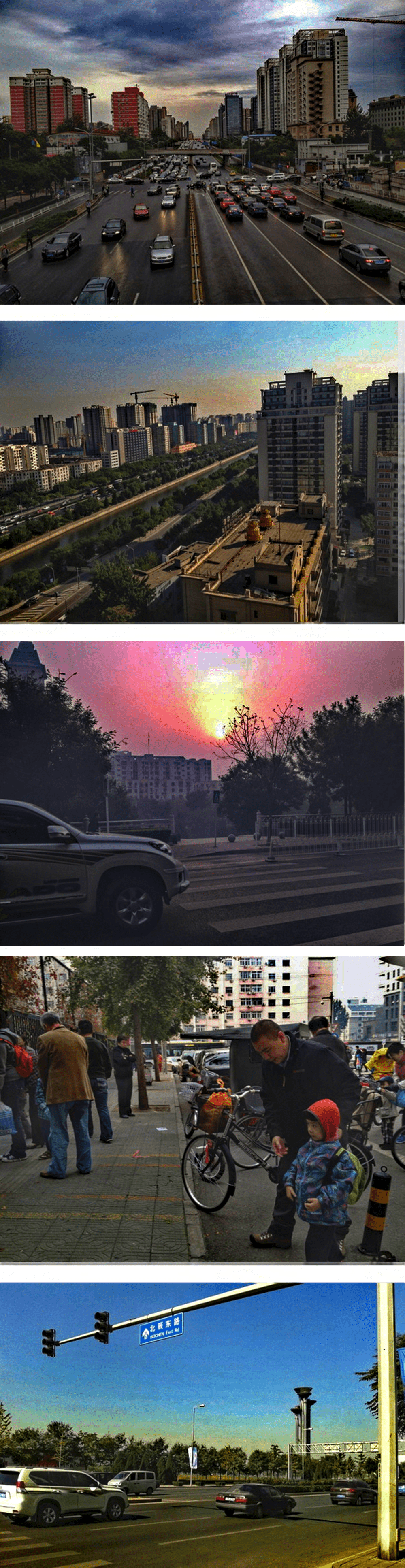}  
}
\subfloat[DehazeNet]{  \centering
  \includegraphics[width=0.08\linewidth]{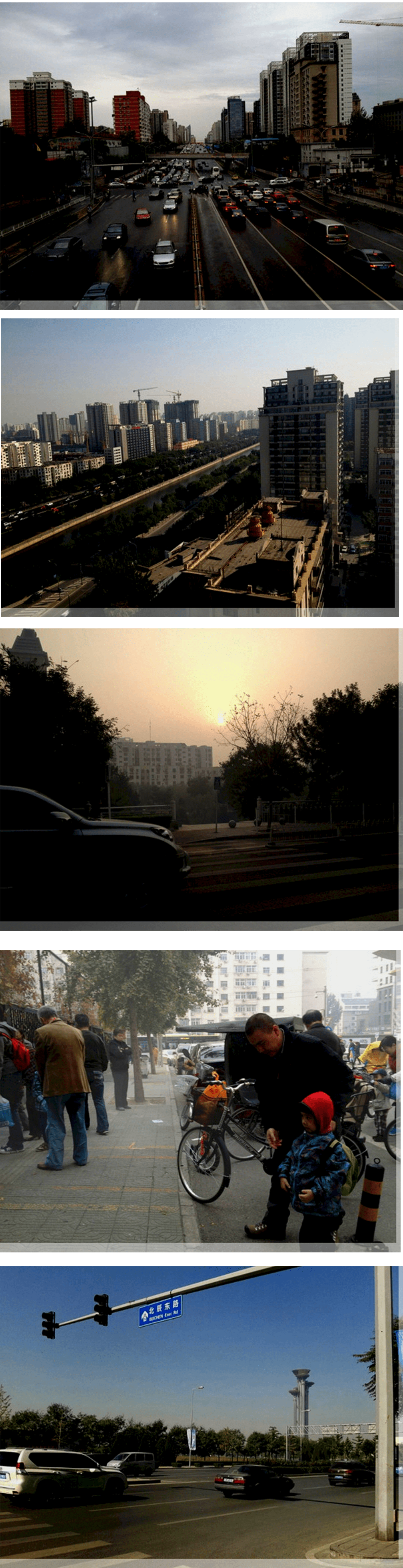}  
}
\subfloat[AOD-Net]{  \centering
  \includegraphics[width=0.08\linewidth]{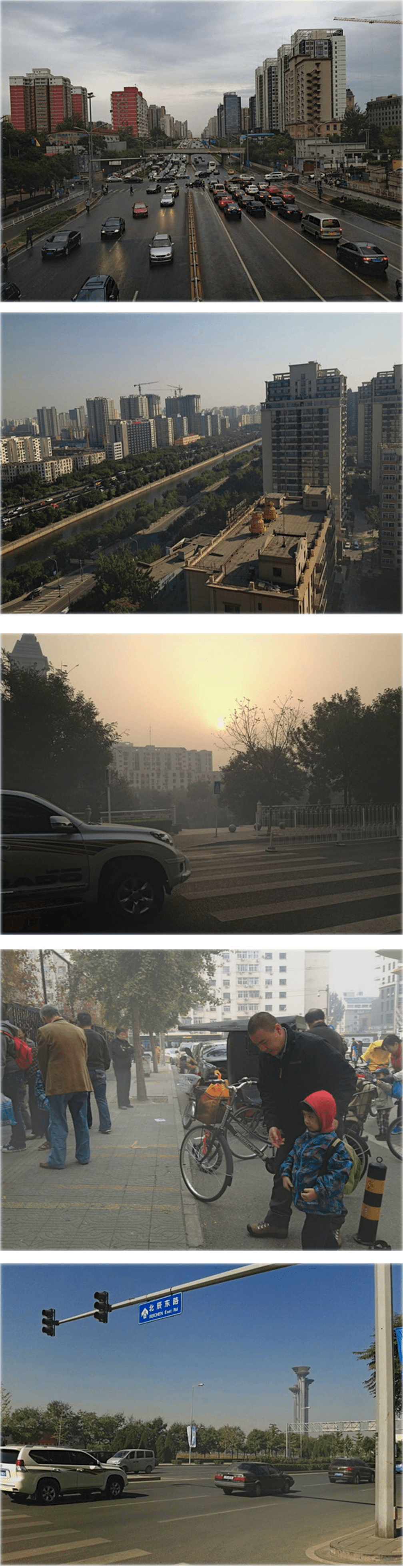}  
}
\subfloat[MSCNN]{  \centering
  \includegraphics[width=0.08\linewidth]{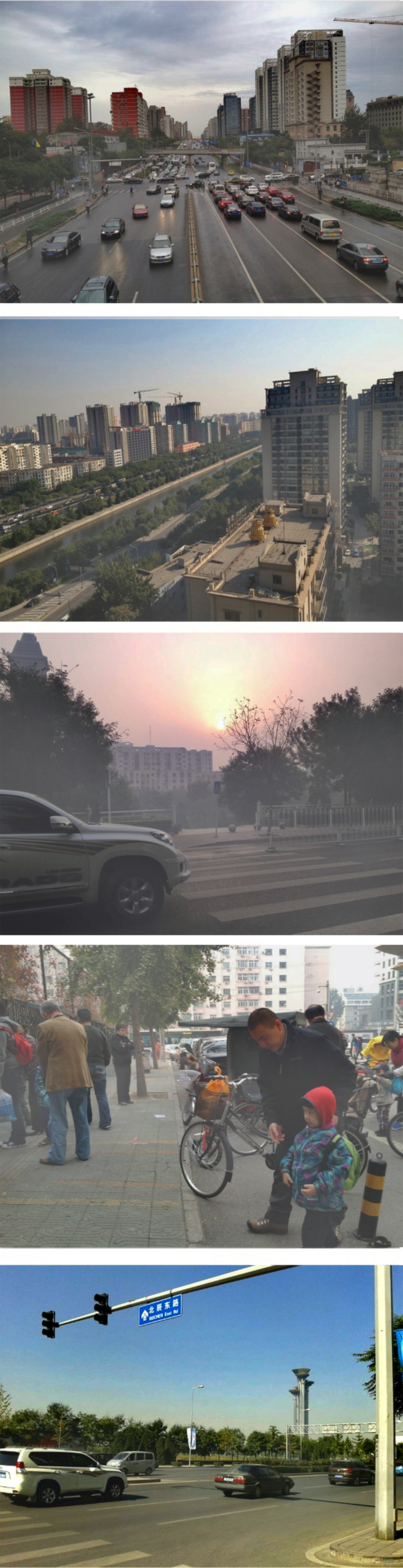}  
}
\subfloat[DCPDN]{  \centering
  \includegraphics[width=0.08\linewidth]{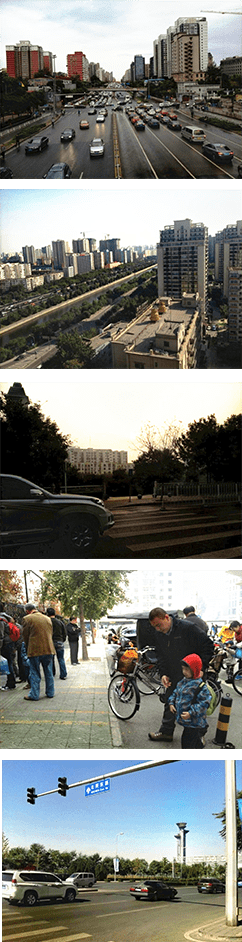}  
}
\subfloat[OTGAN (ours)]{
  \includegraphics[width=0.08\linewidth]{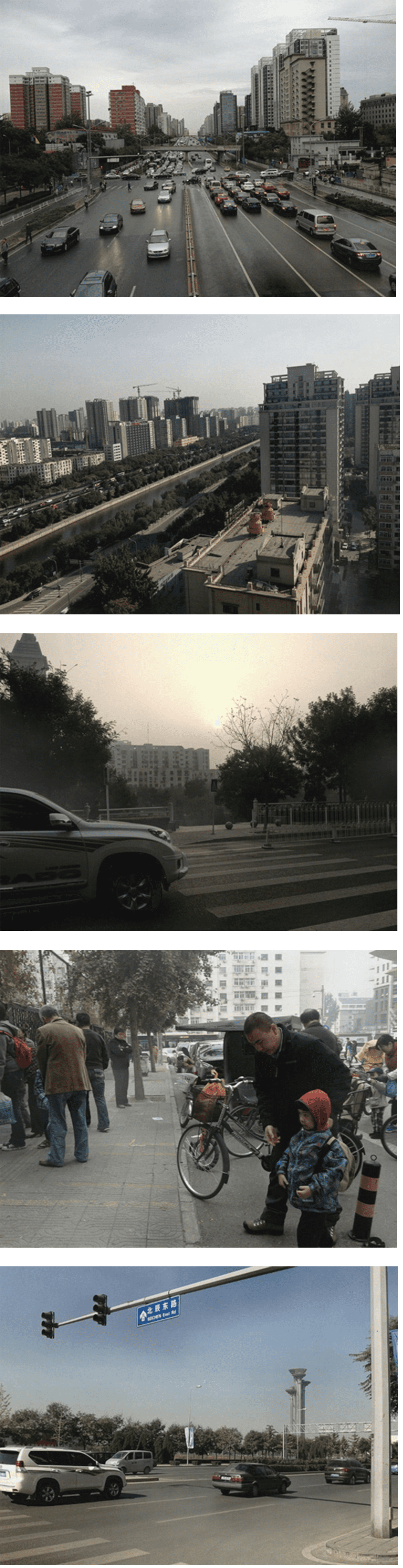}  
}
\subfloat[Ground-truth]{
  \includegraphics[width=0.08\linewidth]{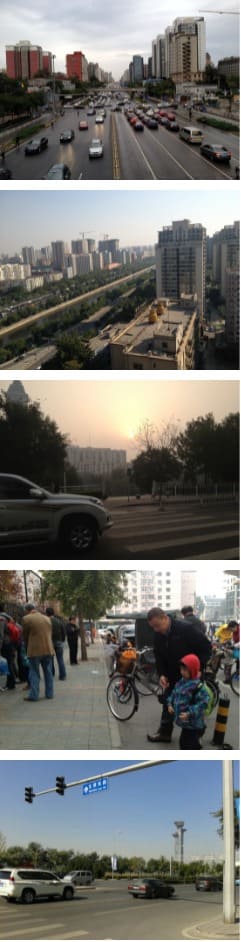}
}}
\caption{Qualitative comparison of various methods on SOTS-outdoor dataset}
\label{qlfig1}
\end{figure*}

\begin{figure*}[!ht]
\captionsetup{justification=centering}
\resizebox{\textwidth}{!}{
\subfloat[Hazy image]{
  \includegraphics[width=0.1\linewidth]{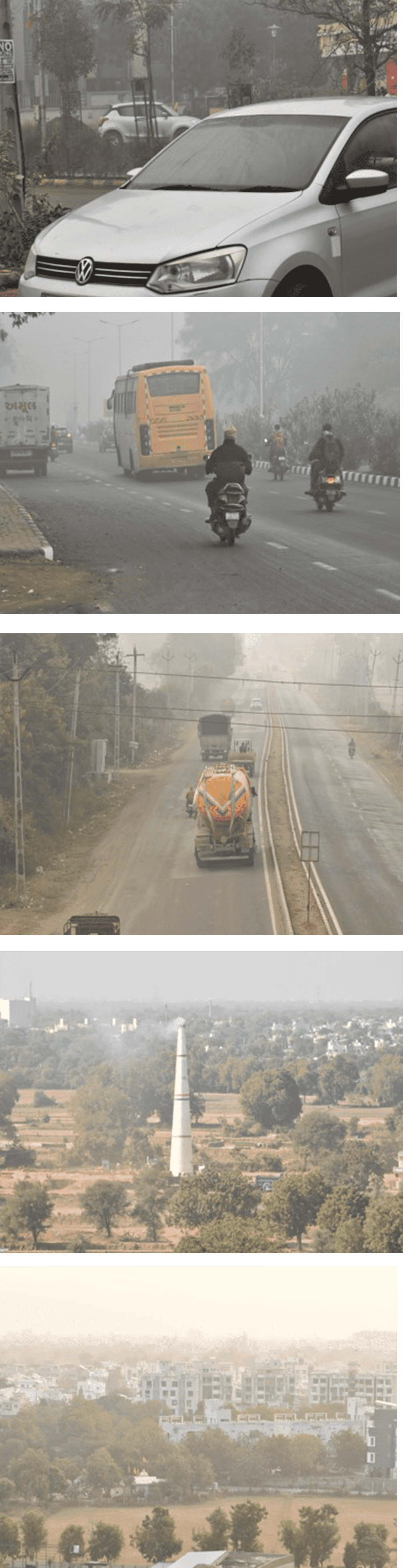}  
}\subfloat[DCP]{
  \includegraphics[width=0.1\linewidth]{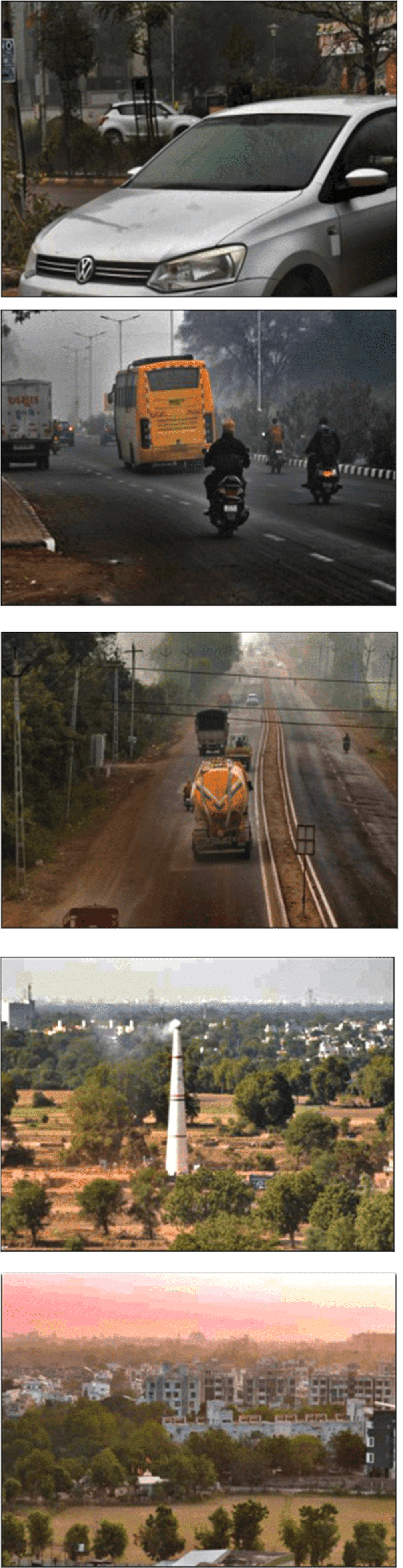}  
}
\subfloat[CAP]{
  \includegraphics[width=0.1\linewidth]{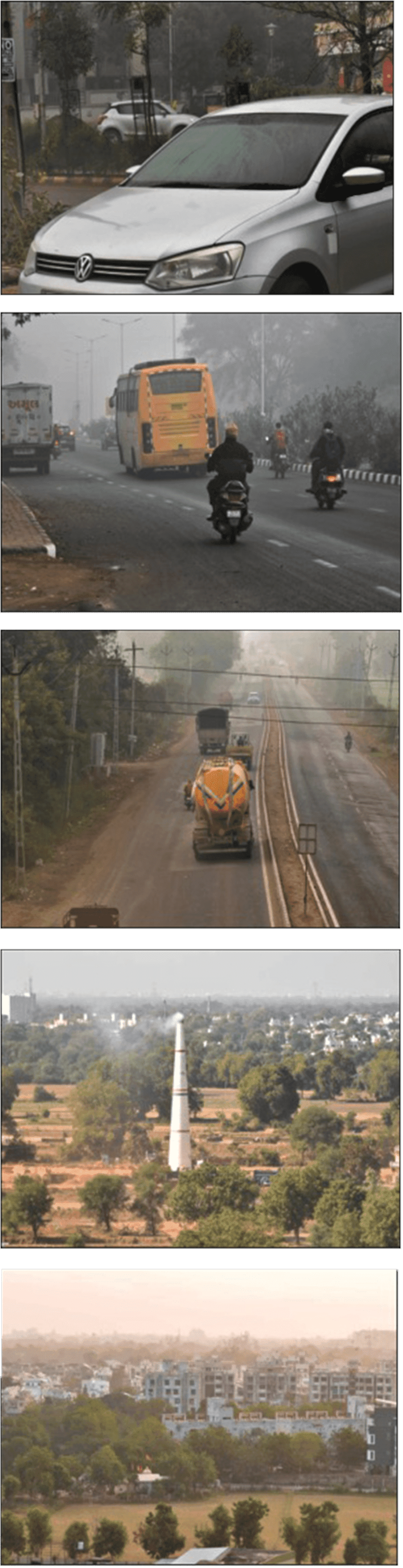}  
}
\subfloat[NLD]{
  \includegraphics[width=0.1\linewidth]{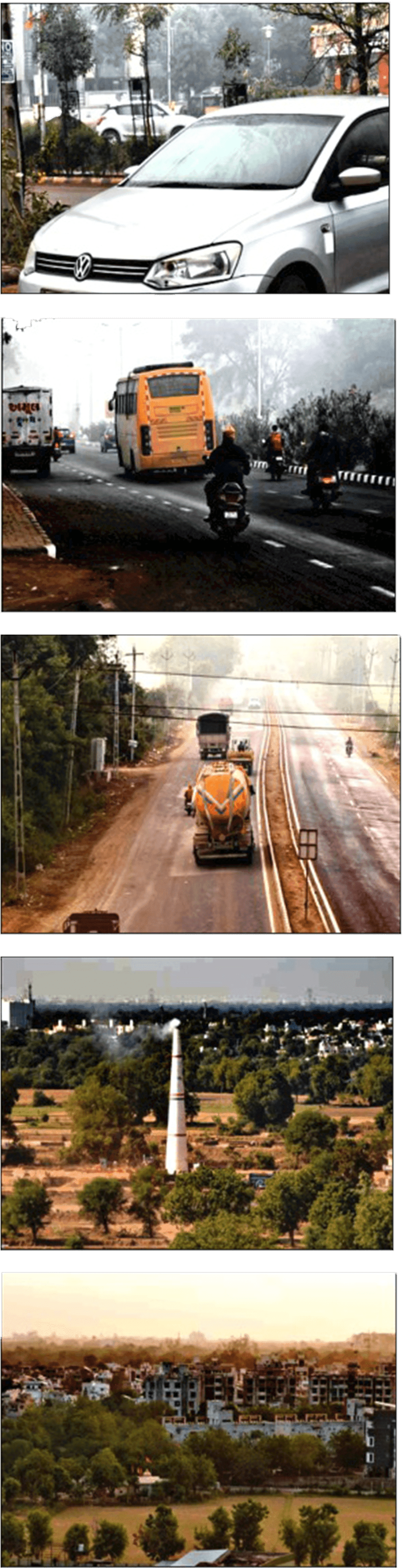}  
}
\subfloat[BCCR]{
  \includegraphics[width=0.1\linewidth]{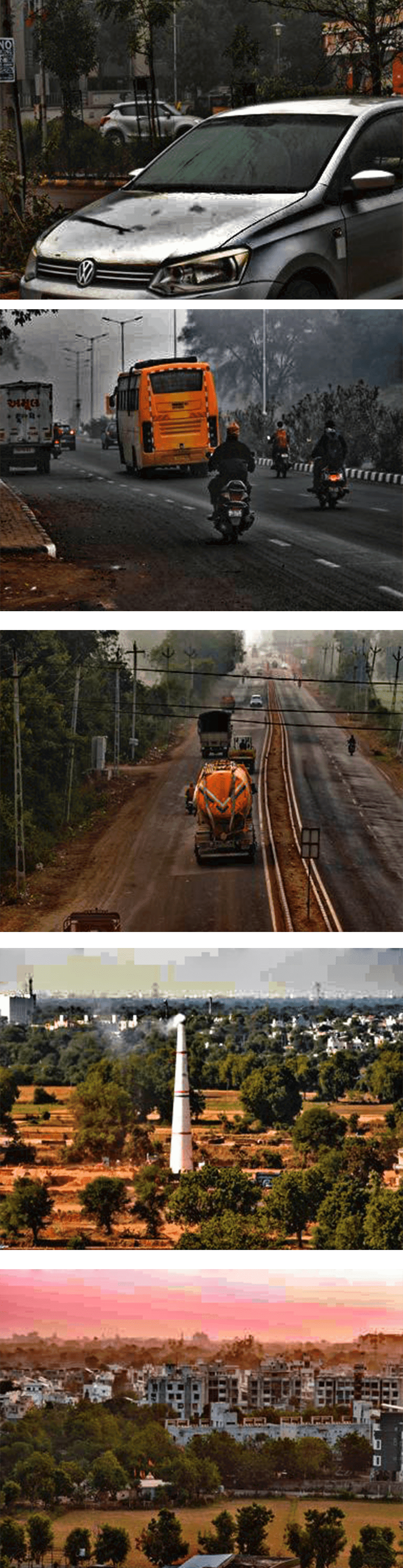}  
}
\subfloat[DehazeNet]{  \centering
  \includegraphics[width=0.1\linewidth]{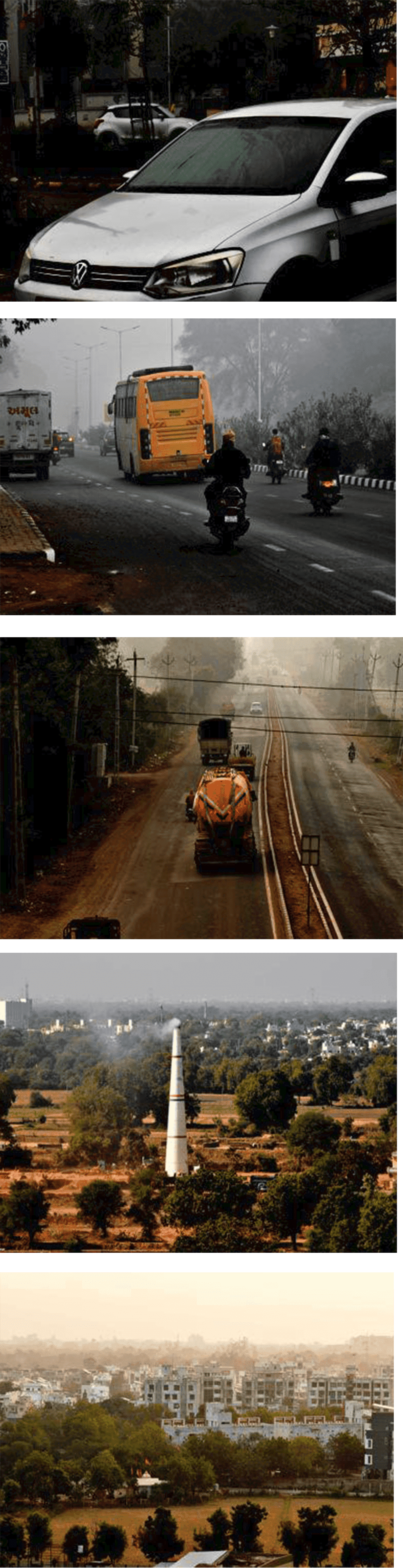}  
}
\subfloat[AOD-Net]{  \centering
  \includegraphics[width=0.1\linewidth]{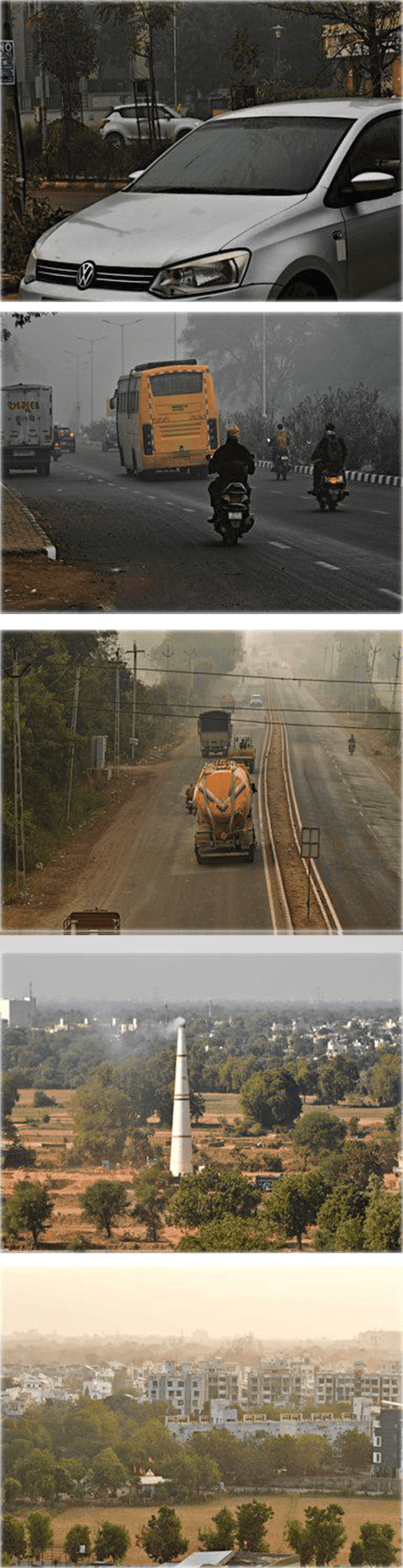}  
}

\subfloat[MSCNN]{  \centering
  \includegraphics[width=0.1\linewidth]{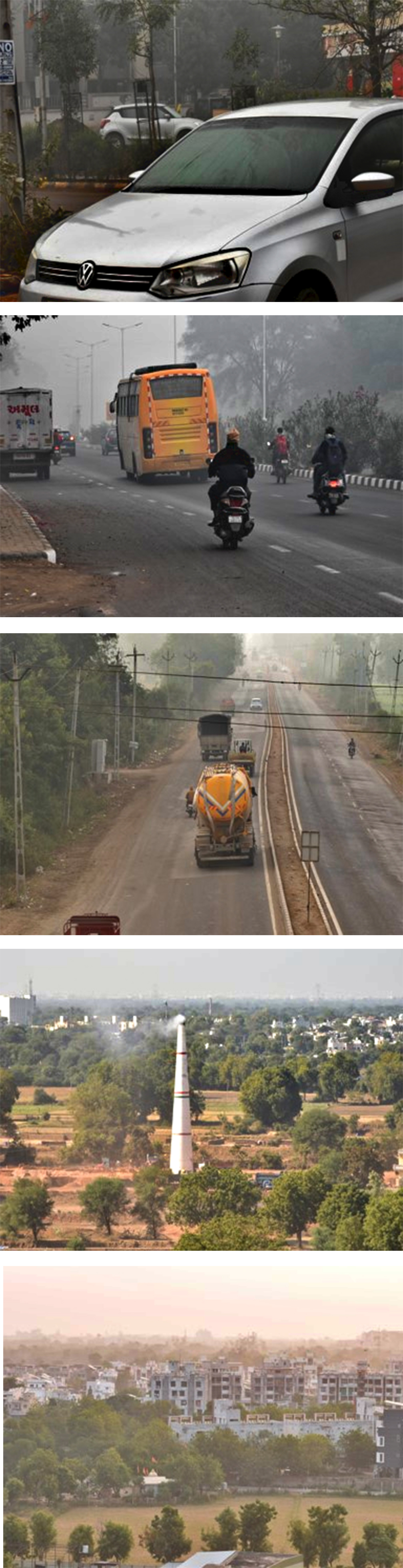}  
}
\subfloat[DCPDN]{  \centering
  \includegraphics[width=0.1\linewidth]{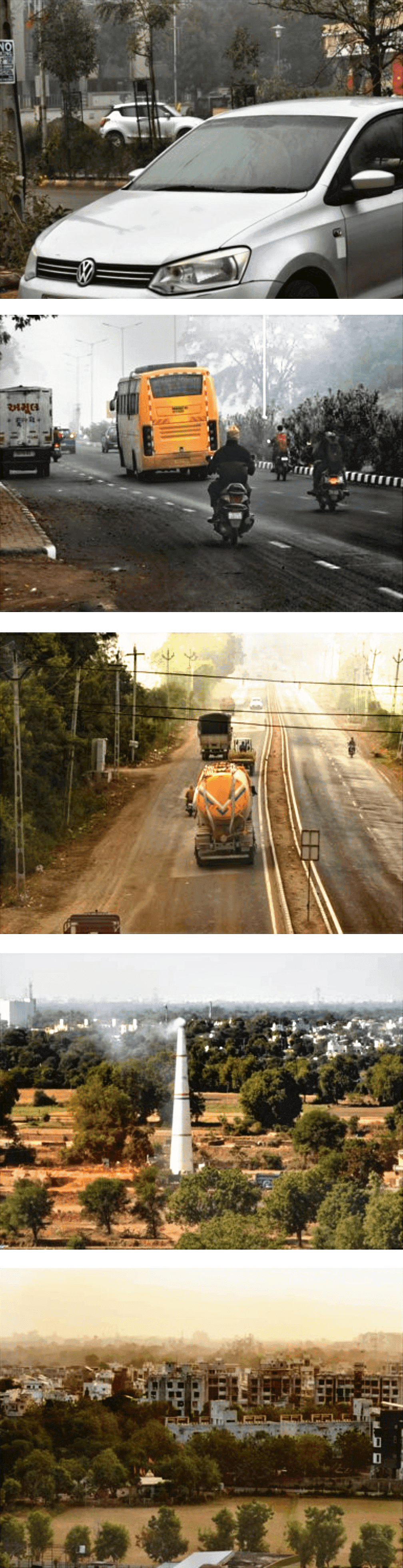}  
}
\subfloat[OTGAN (ours)]{
  \includegraphics[width=0.1\linewidth]{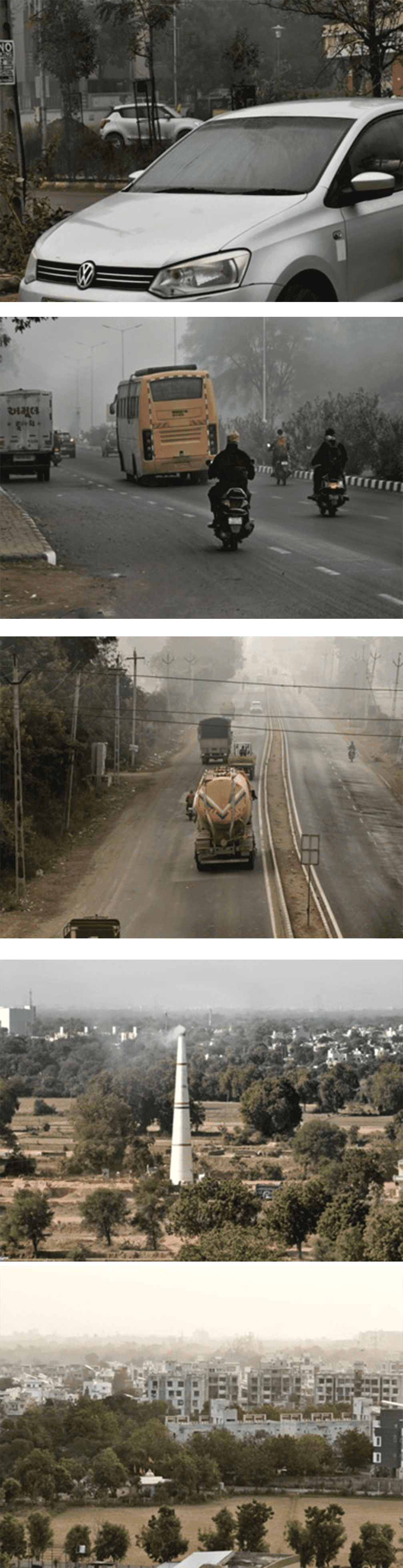}  
}}

\caption{Qualitative comparison of various methods on real-world images}

\label{qlfig2}
\end{figure*}

\begin{table*}[ht]

\caption{Quantitative analysis showing PSNR/SSIM scores (higher the better) for SOTS(Outdoor and Indoor) and HSTS}
\label{qntable}
\resizebox{\textwidth}{!}{
\begin{tabular}{lccccccccccc}

\hline
\hline
              & DCP         & CAP         & BCCR        & NLD        & DehazeNet    & DCPDN       & AOD-NET      & MSCNN       & GFN         &Deep Energy& OTGAN   \\ 
\hline 
SOTS(Outdoor) & 17.55/0.798 & 22.28/0.912 & 15.48/0.782 & 18.05/0.803 & 22.74/0.856 & 19.68/0.882 & 21.34/0.924 & 19.55/0.864 & 21.48/0.837 &          24.08/0.933   &\textbf{25.28/0.935} \\
SOTS(Indoor)  & 20.14/0.871 & 19.06/0.835 & 16.87/0.789 & 17.28/0.748 & 21.14/0.846 & 15.77/0.817 & 19.37/0.850 & 17.12/0.804 & \textbf{22.33/0.879} &  19.25/0.832   & 21.12/0.873\\
HSTS          & 17.21/0.799 & 21.53/0.866 &15.09/0.737 &17.63/0.792 & 24.48/0.916 & 20.40/0.883  & 21.57/0.921 & 18.28/0.842 & 22.93/0.873 &     24.44/\textbf{0.933}          &\textbf{25.42}/0.929    \\
\hline
\hline
\end{tabular}
}
\end{table*}

\begin{figure*}[t]
    \centering
    \subfloat[\label{psnr_split}]{
    \includegraphics[width=0.45\textwidth]{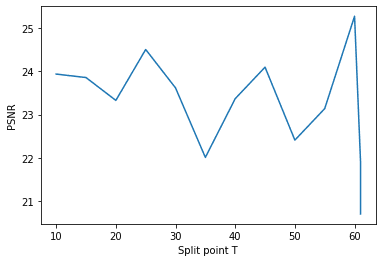}
    }
    \subfloat[\label{ssim_split}]{
    \includegraphics[width=0.45\textwidth]{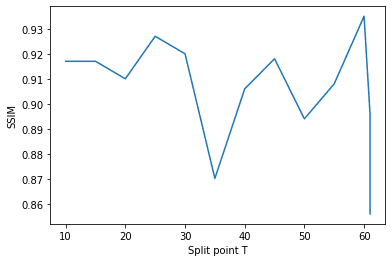}
    }
    \caption{(a) PSNR value for different split points ($T$) (b) SSIM value for different split points ($T$)}
    \label{split_point}
\end{figure*}

\subsection{Qualitative and Quantitative analysis}
 
In this section, the performance of the proposed method is compared with the existing state-of-the-art methods. We have compared our model with DCP \cite{dcp}, CAP \cite{cap}, NLD\cite{nld}, BCCR\cite{bccr}, DehazeNet\cite{dehazenet}, MSCNN\cite{mscnn}, AOD-Net\cite{aod}, DCPDN\cite{dcpdn}, GFN \cite{gfn} and Deep-energy\cite{deepenergy}, where the first four are prior based methods and remaining are learning-based methods. SOTS of RESIDE is used as the testing dataset. Various quality metrics such as Peak Signal-to-Noise Ratio (PSNR), Structural Similarity Index (SSIM)\cite{ssim}, 
and Natural Image Quality Evaluator (NIQE) \cite{niqe} are used as scoring metrics for quantitative analysis of the dehazed images obtained from different methods. Fig. \ref{qlfig1} shows the qualitative comparison of the dehazing methods for SOTS-outdoor dataset along with the ground truth image. It can be observed that all methods are able to remove different amount of haze from the hazy image but the results obtained using the proposed method removes haze to a larger extent and also retains true colors of the image. The dehazing results obtained from BCCR, DCP, NLD methods are over-saturated in terms of the colors and look unrealistic. The results obtained from Dehazenet are darker as compared to our method. This can be seen from third image in sixth column of Fig. \ref{qlfig1},  where the trees present in the image have dark color close to black while the result obtained from our method (tenth column) has true colors and it is easy to distinguish between different objects in the image. The DCP and BCCR overestimate the color of sun and we can observe ringing artifact near the sun while our method does not contain any such artifacts. Moreover, each method produces different color of sky (see fifth row of Fig. \ref{qlfig1}) but the color produced by our method is near to the ground truth image. Table \ref{qntable} shows the quantitative comparison of the methods in terms of PSNR and SSIM scores for SOTS and HSTS datasets.  It can be observed that our proposed method has the highest PSNR and SSIM scores for SOTS outdoor dataset as compared to other methods. The results obtained from our method on indoor dataset are not the highest but they are competitive to other methods like BCCR, NLD, DCPDN and Deep energy.

In order to validate the performance of the proposed architecture on real world images, we have chosen few images from the real world hazy dataset (Fig.\ref{rwdata}) and compared the results of our method with other methods. Fig. \ref{qlfig2} shows the qualitative comparison on real world hazy images. The following observations worth noticeable are as follows: the results obtained by BCCR are darker when compared to other methods. The CAP and MSCNN methods are not able to remove most of the haze from the hazy images. Moreover, it can be concluded that all the methods struggles to dehaze the image completely, but the amount of haze removed by our method is more when compared with other methods. As the real-world images do not have the ground truth image for comparison it is not possible to evaluate PSNR and SSIM values for the real world images. A no-reference image quality metric NIQE \cite{niqe} is used to measure the quality of the dehazed image. It is a no-reference metric that compares the features of the given image with Natural Scene Statistic (NSS) model. This model is constructed using natural and undistorted image corpus. A lower value of NIQE represents a better perceptual quality of the image. Table \ref{niq_table} shows the NIQE score of the proposed method along with the other methods. It can be observed that the proposed method gives the lowest value of NIQE compared to other methods. 

\begin{table}[h]
\centering

\caption{NIQE(LOWER IS BETTER) score on real world dataset}
\label{niq_table}
\begin{tabular}{ccccccc}

\hline
\hline

         & DCP     & CAP    &BCCR & AOD-NET & DCPDN    & OTGAN         \\
\hline
NIQE     & 9.4     & 9.6    &9.7  & 9.5     & 11.9    & \textbf{9.1}         \\\hline
\hline
\end{tabular}
\end{table}

\subsection{Threshold analysis}
As discussed in Section \ref{tar}, the proposed architecture is divided into two branches, one for lower frequencies and the other for higher frequencies. The split point $T$ divides the frequency cube into  two parts $f_{low}$ and $f_{high}$. The architecture is trained with different value of $T$ and the PSNR and SSIM scores are calculated for each of them. Fig. \ref{split_point} shows the PSNR and SSIM  scores for different values of $T$. It can be observed from the figure that at $T = 60$ the average PSNR and SSSIM scores achieves the highest value. Based on this observation we have selected $T = 60$ for our experimental work.
 
\section{Limitation of our Model}

\label{failcase}

\begin{figure}[h]
    \centering
    \subfloat[\label{fail1}]{\includegraphics[width=0.45\linewidth]{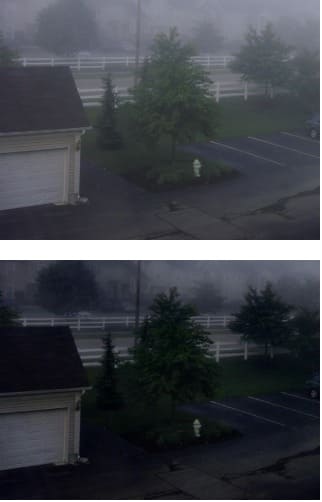}}
    \subfloat[\label{fail2}]{
    \includegraphics[width=0.45\linewidth]{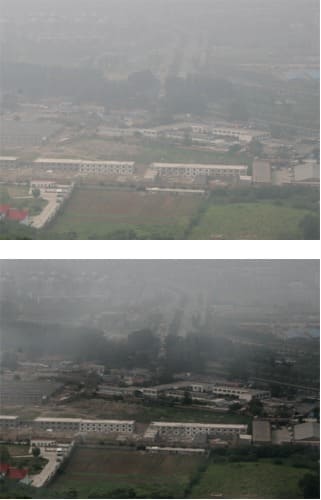}}
    \caption{Failure cases: Upper row shows hazy input. Lower row shows output of our model.}
    \label{fail}
\end{figure}

For the real world images suffering from severely low lighting conditions or dense haze, most existing work fails to produce good results. It has been shown that the proposed work performs better in most of the cases. However, if the above mentioned condition worsen then the performance of the proposed work will also decrease. In particular, when the images with low light conditions is passed to our model, the dehazed image is dark and objects are not clearly visible (refer Fig. \ref{fail1}). The same is true for images with dense haze intensity (refer Fig. \ref{fail2})

\section{Conclusion} \label{conc}
We have proposed a novel end-to-end image dehazing architecture using Orthogonal Transform based Generative Adversarial Network, which performs image dehazing in the Krawtchouk transform domain. The proposed model directly estimates a clear image instead of estimating the transmission map. The Krawtchouk coefficients are used to differentiate between low and high-frequency components of the image which is then utilized to recover the haze-free image from the hazy input image. When compared with existing methods, our proposed method provides competitive results. The visual comparison shows that results obtained from our method look more realistic and recovered clear images with true colors. 

\section{Acknowledgement}
The authors are thankful to Department of Science and Technology-The Gujarat Council on Science and Technology (DST-GUJCOST) for financial assistance.


%





\bibliographystyle{plain}
\bibliography{bibliography.bib}

\end{document}